\documentclass[11pt,a4paper]{article}
\usepackage[utf8]{inputenc}
\usepackage[top=0.5in, bottom=1.0in, right=1.0in, left=1.0in]{geometry}

\title{Commuter Count: \\Inferring Travel Patterns from  Location Data}

\author{Nathan Musoke$^{1,2}$, Emily Kendall$^{1}$, Mateja Gosenca$^{1,3}$,\\ Lillian Guo$^{1}$, Lerh Feng Low$^{1}$, Angela Xue$^{1}$,\\ and Richard Easther$^{1}$}

\date{March 2022}

\usepackage{graphicx}
\usepackage{hyperref}
\usepackage{amsmath}
\usepackage{amssymb}
\usepackage{mathtools}
\usepackage{bm}

\usepackage[capitalise,noabbrev]{cleveref}
\usepackage{sectsty}
\chapterfont{\color{blue}}  % sets colour of chapters
\sectionfont{\color{blue}}  % sets colour of sections
\subsectionfont{\color{blue}}  % sets colour of subsections

\def\N{\ensuremath{\mathbf{N}}}

\def\M{\ensuremath{\mathbf{M}}}
\def\Mtii{\ensuremath{{M}_{tii}}}
\def\Mtij{\ensuremath{{M}_{tij}}}
\def\Mtji{\ensuremath{{M}_{tji}}}

\def\thetab{\ensuremath{\bm{\theta}}}
\def\thetaij{\ensuremath{\theta_{ij}}}

\def\pivec{\boldsymbol{\pi}}

\def\svec{\ensuremath{\mathbf{s}}}

\usepackage{color}

\allowdisplaybreaks%

\begin{document}

\makeatletter
\begin{titlepage}
        
        \includegraphics[trim={5cm 5cm 5cm 2cm}, height=10cm, angle =270]{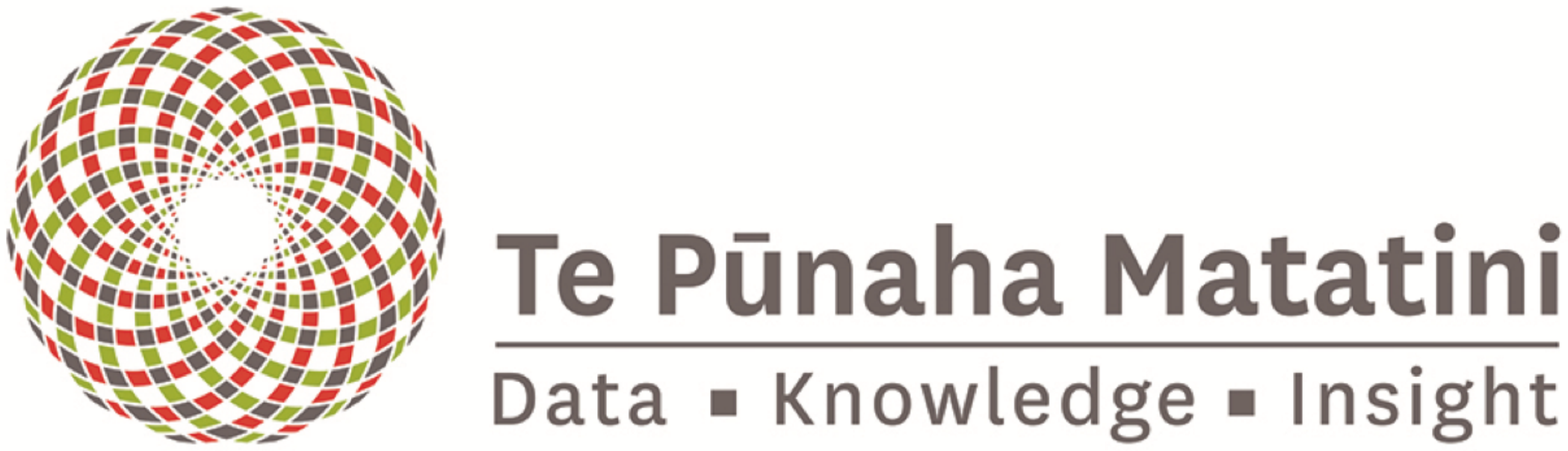}
        \vspace{2cm}
        
        \begin{center}  
            {\huge\color{red}{\@title}\unskip\strut\par}
            \vspace{2cm}
            {\Large\itshape\@author\unskip\strut\par}
            \vspace{1cm}
            $^1$Department of Physics, University of Auckland, New Zealand\\
            
            $^2$Department of Physics and Astronomy, University of New Hampshire, USA\\

            $^3$Faculty of Physics, University of Vienna, Boltzmanngasse 5, 1090 Vienna, Austria\\

             \vspace{1cm}
                   
             For correspondence \href{mailto:emily.kendall@auckland.ac.nz}{emily.kendall@auckland.ac.nz} 	     and  \href{mailto:r.easther@auckland.ac.nz}{r.easther@auckland.ac.nz}
        \end{center}    
        
        \vfill
        \includegraphics[height=2.3cm]{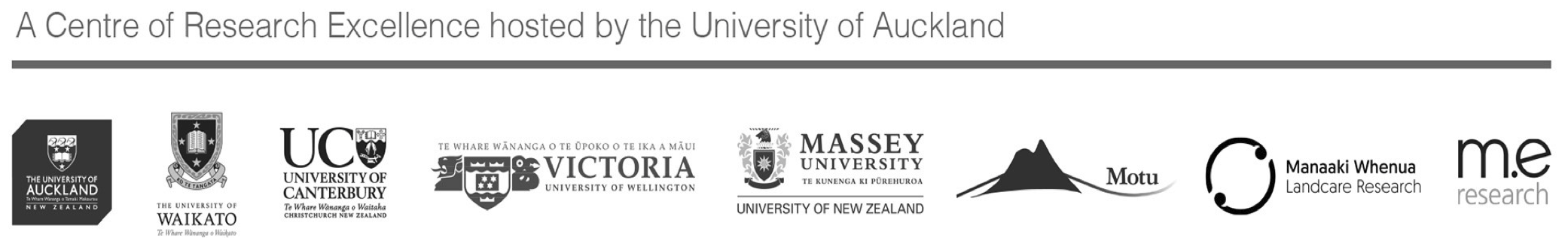}
         
\end{titlepage}
\makeatother

\section{Introduction}%
\label{sec:intro}

The movement of people between geographical regions and their personal interactions are key determinants of the spread of pathogens such as Covid-19.
While interpersonal connections occur on scales ranging from individual households to international travel, interactions between people in the course of their daily routines provide a key ``meso'' layer in any detailed analysis of pathogenic transmission.

The accumulation and analysis of data on the daily activities of individuals has privacy implications and commercial sensitivities, creating (entirely legitimate) barriers to its use by modelers. However, while it is unlikely that detailed trajectories of individuals through the course of a day will be shared outside of tightly controlled environments, aggregated spatio-temporal data can often be made available.

In this Working Paper we analyse strategies for using aggregated spatio-temporal population data acquired from telecommunications networks to infer travel and movement patterns within regions. Specifically, we focus on hour-by-hour cellphone counts for the SA-2 geographical regions covering the whole of New Zealand~\cite{SA2} and base our work on algorithms described by Akagi {\it et al.\/}~\cite{akagi_fast_2018}. This Working Paper describes the implementation of these algorithms, their ability to yield inferences based on this data to build a model of travel patterns during the day, and lays out opportunities for future development.

Our testing data set consists of cellphone counts during January and February 2019 and 2020, where counts are given for individual New Zealand SA-2 geographical regions on an hourly basis. For reference, there are 2253 SA-2 regions in New Zealand. The regions vary in area, such that their typical population is on the order of a few thousand people.
The Greater Auckland region contains approximately 600 SA-2s whereas in remote parts of the South Island a similar geographical area might correspond to just handful of SA-2s.\footnote{There are several exceptions, including offshore islands, which are very thinly populated. This approach also implicitly assumes that cellphone counts are a good proxy for the locations of the majority of the population.}

We focus on the two algorithms,  developed by Akagi and colleagues,  referred to as the `exact' and `approximate' methods. These algorithms use hour-by-hour population counts to estimate bidirectional flows between pairs of geographical regions. Long-distance travel over short time periods is penalised by a simple function of the physical separation between regions. Furthermore, a strict upper bound can be applied to the distance plausibly travelled between successive time steps, so that not all possible region pairings are viable.  The algorithms adapt naturally to ``real'' geographies with complex shapes (rather than a grid-based geometry) and data in which the total number of individuals is not constant, due to phones being turned on or off or moving in and out of coverage areas.

The motivation for this work was to facilitate analyses of the spread of Covid-19\ in New Zealand. However, the treatment here can be applied to any number of tasks requiring models of population flows. This work investigates these algorithms and extends our understanding of their properties and limitations.

Having implemented both the exact and approximate algorithms, we test the consistency of their outputs and find limitations and sensitivities to input parameters which are common to both algorithms. We also identify and address issues that arise when the number of people leaving a given region is roughly similar to the number of destinations available to them, so that the expected number of people moving between many pairs is less than one.
In addition we have developed a simulator which generates synthetic data that allows the algorithms to be explored without access to cellphone counts and the underlying individual trajectories, facilitating additional verification strategies.

Our implementation of the exact algorithm is computationally efficient; far more so than originally reported. In particular,  we can ``solve'' for the Greater Auckland and Waikato regions (encompassing the three Auckland area District Health catchments) in tens of seconds of walltime on a laptop.

This Working Paper is structured as follows.
\Cref{sec:data} provides a quick quantitative survey of the cellphone count data utilised in the inference process. \cref{sec:likelihood} discusses the construction of a likelihood function characterising the probabilities of transitions between regions and \cref{sec:algorithm} summarises the exact Akagi algorithm to maximise this likelihood, describes its implementation in a computationally efficient Python code\footnote{Available at \url{https://github.com/auckland-cosmo/FlowStock}}, and identifies issues with its application to this problem. \Cref{sec:simulated_data} introduces a simple data simulator used to validate the code, while \Cref{sec:validation_of_method_and_implementation} looks at the sensitivity of the results to key parameters in the model.
\Cref{sec:approx_inference} describes the alternative approximate algorithm  proposed by Akagi \textit{et al.} and contrasts its output to the  exact algorithm. In \cref{sec:south_alt} we sketch an approach to compare our results to external sources of commuter information.
Finally, \Cref{sec:summary} provides a brief summary of our experiences and identifies areas for further exploration. We discuss possible improvements and extensions to the algorithms, and highlight issues with these algorithms that might limit their utility.

\section{Cellphone Count Data }%
\label{sec:data}

\begin{figure}[tb]
    \center{\includegraphics[width=\textwidth, trim={0 0 0 0.5cm}]{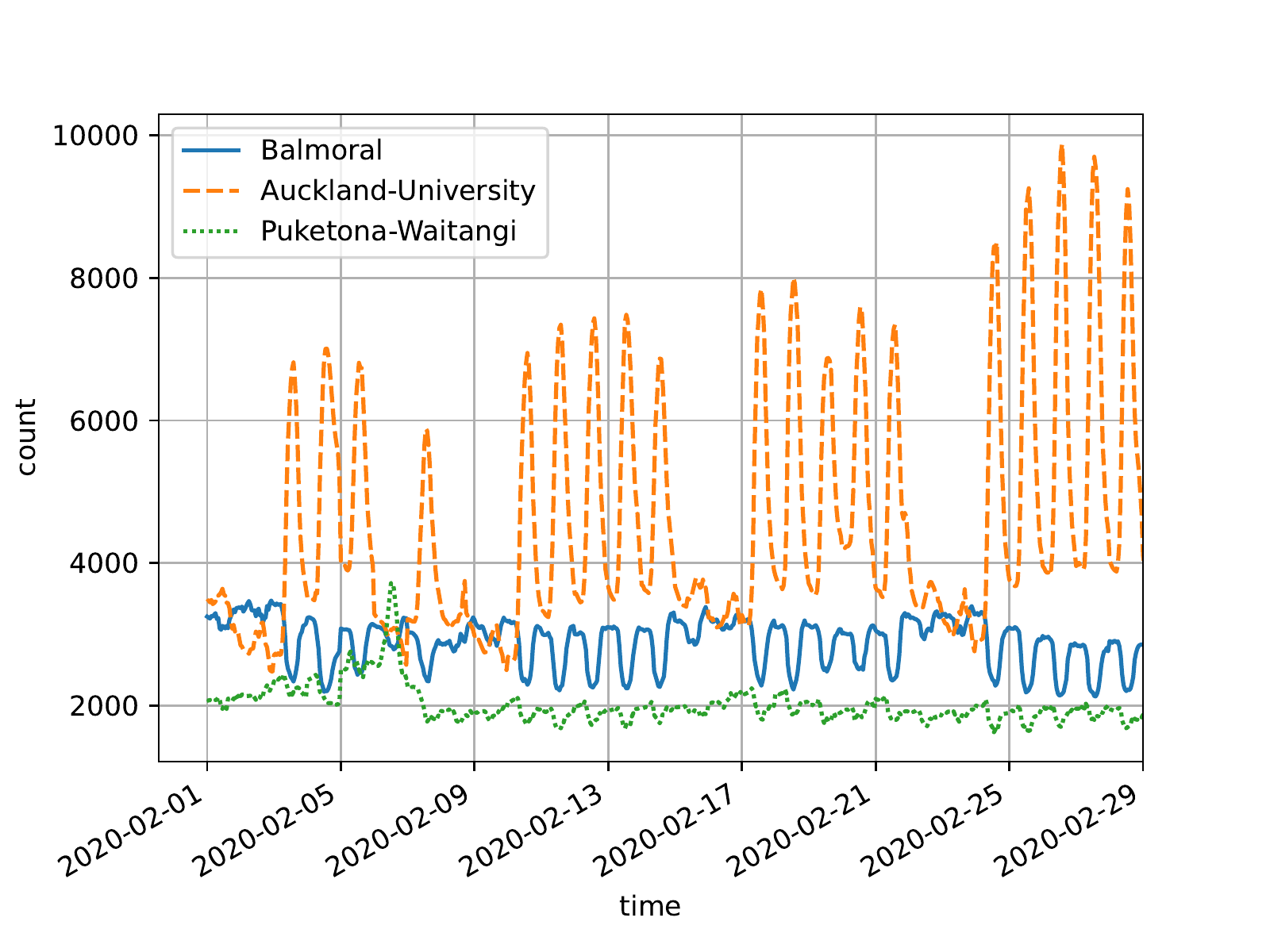}}
    \caption{%
        Representative counts throughout a month.
        There are clearly large daily commutes in and out of the central-city SA-2 region \texttt{Auckland-University}, anti-correlated with flows to the residential area \texttt{Balmoral}.
        There is a discernible difference between workday and weekend counts.
        Inspecting data from \texttt{Puketona-Waitangi}, containing the Waitangi Treaty grounds, one can see a significant increase in the lead up to February 6th. \label{fig:rep_data}
    }
\end{figure}

\begin{figure}[tb]
    \center{\includegraphics[width=\textwidth]{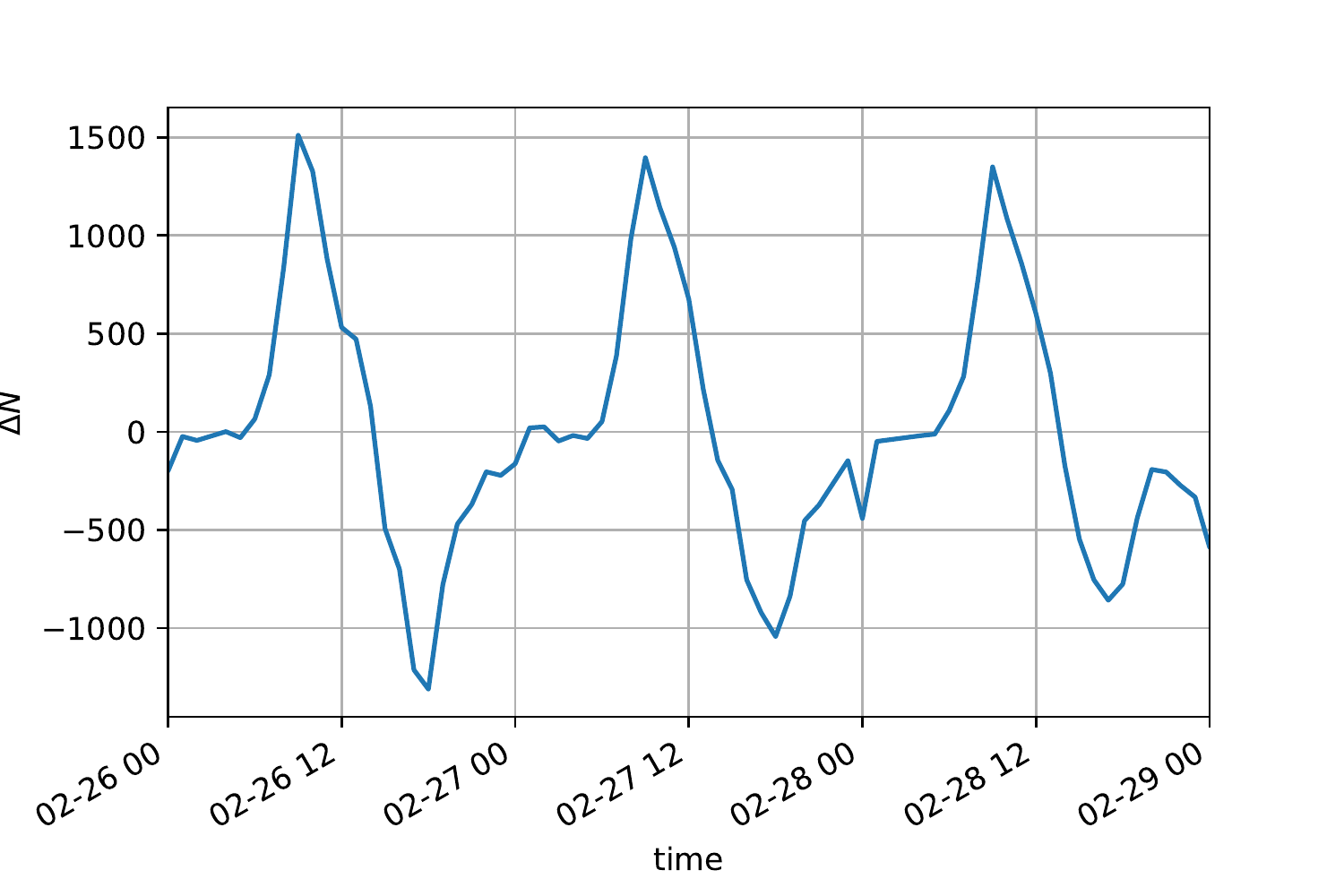}}
    \caption{%
        Hourly differences in the count in the \texttt{Auckland-University} area during the 26th to 28th of February 2020.
        One can see a sharp morning rush hour and less pronounced evening rush hour.
        There is an anomaly at midnight on the 28th.
        Such features are common at midnight and appear to be artifacts associated with the capture and processing of the data by the telecommunications providers.
            } \label{fig:delta}
\end{figure}

Our analysis uses aggregated cellphone count data gathered from New Zealand telecommunications companies. In particular, this data gives hourly counts of the number of cellphones active in each SA-2 region. Note that the term `active' applies to all cell phones which are powered on and detectable by the cell network; a cell phone does not need to be in use to be considered active. Within this data, it is possible to clearly discern  patterns in population flow, for example during weekends, holidays, or large gatherings.
\Cref{fig:rep_data} provides some representative examples.\footnote{February 6th is a public holiday in New Zealand, during which there is often a large gathering at Puketona-Waitangi to commemorate the signing of the Treaty of Waitangi.}

It should be noted that each cell phone is counted in only one SA-2 region per hour. This is reflected by the conservation of the total cell phone count over time. Indeed, while a cell phone may be in range of multiple cell towers at any given moment, it will only use a single tower for communication at any one time, as determined by the relative signal strength. Hence, when the instantaneous count is performed on an hourly basis, each cell phone is associated with only one tower/SA-2 region. As the hourly data represents instantaneous counts, movements between SA-2 regions/cell towers on timescales smaller than one hour are not captured. 

While most of the adult population is likely to carry a cellphone with them on a day-to-day basis, there is no guarantee that cellphone counts map uniquely to individuals or that the mapping is unbiased. Indeed, we expect that certain demographics --- e.g.\ the very young or very old, and the economically deprived --- may be missed in this data. Furthermore, populations with 0 or multiple phones will be heavily biased in age, social class, and other areas that are correlated with infection risk factors. Unfortunately, the currently available data on cell phone access is not sufficiently detailed to incorporate into our modelling at this time. While some relevant 2018 census data is available \cite{cell_access}, it only provides information on access to telecommunication systems at the household, rather than individual, level. Furthermore, the census data includes no information for 7.7\% of households. While a detailed study of cell phone ownership is outside of the scope of this work, it is expected that data from future national surveys may improve our ability to correlate the movements of cell phones with the movements of individual persons. 

Finally, we also note that the data exhibits frequent discontinuities in regional counts at midnight, when cell tower data is apparently ``rebaselined'', as shown in \Cref{fig:delta}. However, since our focus will be on movements during the working day this is of limited relevance to our analysis.

\section{Log-Likelihood and Likelihood Gradient}
\label{sec:likelihood}

\begin{table}
    \centering
    \label{tab:symbols}
    \begin{tabular}{c p{10cm}}
        $V$ & set of regions
        \\
        $n$ & number of regions, $|V|$
        \\
        $T$ & number of snapshots
        \\
        $\N$ & $n \times T$ matrix of counts in regions at each snapshot
        \\
        $\mathbf{d}$ & $n \times n$ matrix of distances $d_{ij}$ from region $i$ to $j$
        \\
        $K$ & distance cutoff
        \\
        $V$ & set of regions
        \\
        $\Gamma_i$ & set of neighbours of region $i$; $\{j \in V | d_{ij} \leq K\}$
        \\
        $\Mtij$ & the number of people who move from $i$ to $j$ between $t$ and $t+1$; $\M$ is a $(T-1) \times n \times n$ array
        \\
        $\pi_i$ & departure probability of region $i$
        \\
        $s_i$ & gathering scores for region $i$
        \\
        $\beta$ & scalar distance weighting
        \\
        $\theta_{ij}$ & probability for a person to move from region $i$ to $j$ between snapshots
        \\
        $\mathcal{C}(\M; \N)$ & cost function to enforce number conservation
        \\
        $\lambda$ & weighting of cost function
        \\
        $\mathcal{L}(\M, \pivec, \svec, \beta; \N, \lambda, \mathbf{d}, K)$ & likelihood of $\M$, $\pivec$, $\svec$, and $\beta$ given data $\mathbf{N}$ and assumptions $\lambda$, $\mathbf{d}$, $K$
        \\
        $\epsilon$ & convergence threshold for iterative optimisation
    \end{tabular}
    \caption{%
        Symbols used in the text.
        Bold symbols are non-scalar quantities.
    }
\end{table}

Following Akagi \textit{et al.}, we introduce a probability of transition between different regions,
\begin{equation}
    \label{eq:prob}
    P(\M | \N,\thetab)
    =
    \sum_{t=0}^{T-2}\sum_{i\in V} \left(
        \frac{N_{ti}!}{\prod_{j\in \Gamma_i}\Mtij!} \prod_{j\in \Gamma_i} \theta_{ij}^{\Mtij}
    \right).
\end{equation}
Here $N_{ti}$ denotes the observed number of people in region $i$ at step $t$ (the algorithms can consider multiple time slices), which is provided as  input data. The number of transitions from region $i$ to $j$ at step $t$ is represented by $M_{tij}$. The $M_{tij}$  are the quantities we seek to estimate\footnote{Note that $T$ represents the total number of time \textit{slices}, such that there are $T-1$ time \textit{steps} between the slices, labelled from $t=0$ to $t=T-2$.}.
For each starting region $i$, the set of possible destination regions is denoted by
\begin{equation}
    \label{eq:gamma}
    \Gamma_i = \{j \in V | d_{ij} \leq K\}
\end{equation}
where $d_{ij}$ is the distance from region $i$ to region $j$; $K$ is a cutoff distance beyond which people are assumed not to travel in a single time step.\footnote{We assume that the distance metric $d$ corresponds to the centroid-to-centroid distance between SA-2 regions. Centroid coordinates are available at \url{https://datafinder.stats.govt.nz/layer/93620-statistical-area-2-2018-centroid-true/}.}
The probability of a person in region $i$ at time $t$ moving to region $j$ at time $t+1$ is then $\theta_{ij}$. In general, this probability will be dependent on the time of day. For example, commuter traffic into and out of central business districts tends to reverse from morning to evening. It is therefore important that the estimation algorithm be applied across time periods in which transition probabilities may be assumed to be roughly constant.  

The algorithm requires an assumption for the transition probability, which is taken to be
\begin{equation}
    \label{eq:transition_probability}
    \theta_{ij}
    =
    \begin{dcases}
        1 - \pi_i &\text{\ if } i = j\
        \\
        \pi_i \dfrac{s_j \exp(-\beta d_{ij})}{\sum_{k \in \Gamma_i \setminus \{i\}} s_k \exp(-\beta d_{ik})} &\text{\ if } i \ne j
    \end{dcases}
    \,,
\end{equation}

where the $\pi_i$ are components of the vector $\pivec$ of length $n$ which describes the probability of a person leaving their current region. Their possible destinations are weighted by another $n$-vector, $\svec$, where $s_j$ describes the tendency for people to gather in region $j$. For example, regions within the central business district would be expected to have a strong tendency to attract commuters during the morning rush hour. Following Akagi {\textit{et al.}}, we include an exponential penalty on long-distance travel, but note that other forms of penalty are possible.\footnote{As we see below, $\beta$ is one of the parameters we adjust to optimise the fit. In some cases the optimal value of $\beta$ was negative, but often for unrealistically small regions --- and there are also more possible pairings at greater distances. We experimented with the choice $e^{-\beta_1 d_{ij} + \beta_2 d_{ij}^2 }$ but did not pursue it in detail.}
Finally, note that $\svec$ has an arbitrary overall normalisation.

Akagi \textit{et al.} obtain a log-likelihood function from \cref{eq:prob},
\begin{equation}
    \label{eq:L}
        {\cal{L}}'
        =
        \mathcal{L}_0 + \mathcal{L}_1 + \mathcal{L}_2
        - \frac{\lambda}{2}{\cal{C}({\M, \N})},
\end{equation}
where the individual components are given by:
\begin{gather}
    \label{eq:term0}
    \mathcal{L}_0 = \sum_{t=0}^{T-2} \sum_{i} \log(1-\pi_i)  \Mtii,
    \\[2.0ex]
    \label{eq:term1}
    \mathcal{L}_1 = \sum_t \sum_i  \sum_{j \in \Gamma_i \backslash \{i\}} \left( \log(\pi_i ) + \log(s_j) - \beta d_{ij} - \log\sum_{k \in \Gamma_i \backslash \{i\}} s_k e^{-\beta d_{ik}}\right) \Mtij,
    \\[2.0ex]
    \label{eq:term2}
    \mathcal{L}_2 = \sum_t \sum_i \sum_{j \in \Gamma_i } (1 - \log\Mtij) \Mtij,
    \\[2.0ex]
    \label{eq:cost}
    {\cal{C}({\M, \N})}
    =
    \sum_{t=0}^{T-2} \sum_{i} {\left( N_{ti} - \sum_{j} \Mtij\right)}^2
    +
    \sum_{t=0}^{T-2} \sum_{i} {\left( N_{t+1,i} - \sum_{j} \Mtji\right)}^2.
\end{gather}
Stirling's approximation for factorials is used in the course of this derivation; we will revisit this choice in \cref{sub:scaling}.
The diagonal component of the $t$-th transition matrix $M_{tii}$ corresponds to the population that does not leave block $i$ at step $t$.
The cost function $\mathcal{C}(\M, \N)$ is a soft enforcement of number conservation and this is the only place where the overall size of the population enters the likelihood, rather than dimensionless transition rates.
The strength of the cost function is controlled by the parameter $\lambda$.

We estimate flows by maximizing $\mathcal{L}$ with respect to the $n^2$ components of $\M$ (per time step), the $n$ components of $\pivec$ and $\svec$, and the scalar $\beta$.
The distance cutoff can fix some components of $\M$ to zero but this will not always result in a meaningful simplification of the optimisation problem.
For instance, the Auckland region directly couples in excess of 500 SA-2 blocks, creating approximately 250,000 pairs, each of which has a corresponding $M_{tij}$.
Consequently, the application of this algorithm to a  realistic  problem  involves estimating values for $10^5$ to $10^6$ variables.

We perform the optimisation with the SciPy~\cite{LBFGS,Virtanen:2019joe} implementation of the L-BFGS-B algorithm. By default,  derivatives of the target function are evaluated  via differencing, requiring multiple evaluations of the likelihood. Since the complexity of the likelihood and the number of free parameters both grow with the number of possible pairs the optimisation quickly becomes numerically challenging. However, we can greatly improve performance by supplying analytic derivatives as the $\Mtij$ do not appear in complicated combinations within the likelihood.

After some calculation, we find that the derivatives of the terms in \cref{eq:L} are
\begin{gather}
    \label{eq:der_L_0}
    \frac{\partial \mathcal{L}_0}{\partial \Mtij}
    =
    \begin{cases}
        \log(1-\pi_i)
        & i = j
        \\
        0
        & i \ne j
    \end{cases}
    \\[2.0ex]
    \label{eq:der_L_1}
    \frac{\partial \mathcal{L}_1}{\partial \Mtij}
    =
    \begin{cases}
        0
        & i = j
        \\
        \log(\pi_i ) + \log(s_j) - \beta d_{ij} - \log\sum_{k \in \Gamma_i \backslash i} s_k e^{-\beta d_{ik}}
        & j \in \Gamma_i \setminus \{i\}
        \\
        0
        & j \notin \Gamma_i
    \end{cases}
    \\[2.0ex]
    \label{eq:der_L_2}
    \frac{\partial \mathcal{L}_2}{\partial \Mtij}
    =
    \begin{cases}
        - \log\Mtij
        & j \in \Gamma_i
        \\
        0
        & j \notin \Gamma_i
    \end{cases}
    \\[2.0ex]
    \label{eq:der_C}
    \frac{\partial \mathcal{C}}{\partial \Mtij}
    =
    - 2 \left(N_{ti} - \sum_{l} M_{til}\right)
    - 2 \left(N_{t+1,j} - \sum_{l} M_{tlj}\right)
\end{gather}
While computing the likelihood requires summing $\mathcal{O}(n^2)$ terms, the derivative of the cost function requires summing $\mathcal{O}(n)$ terms, and each of other terms in the derivative is a {\em single\/} term from the sums in $\mathcal{L}$.
Consequently, evaluating the derivative of $\mathcal{L}$ with respect to each of the $n^2$ components of $\M$ will involve $\mathcal{O}(n \times n^2)$ operations whereas approximating them numerically would involve $n^2$ evaluations of the likelihood, for a total cost of $\mathcal{O}(n^2 \times n^2)$.

We may further improve computational efficiency in evaluating both $\mathcal{L}$ and $\partial \mathcal{L}/\partial \Mtij$: when optimising with respect to $\mathbf{M}$, the $\log$ in \cref{eq:term0} and bracketed term in \cref{eq:term1} do not change, and can  be precomputed, offering a significant improvement in efficiency over millions of evaluations.

\section{``Exact'' maximisation algorithm}%
\label{sec:algorithm}

The ``exact'' maximisation algorithm described by Akagi {\textit et al.} requires an iterative maximisation, looping over three separate maximisations until the relative difference in $\mathcal{L}$ changes by less than $\epsilon$, an adjustable parameter. We have implemented the exact algorithm as follows:

\begin{enumerate}
    \item
        Initialise $\M$, $\pivec$, $\svec$, $\beta$.

        This step is unspecified in~\cite{akagi_fast_2018}, but the way it is done has a significant impact on the results of the algorithm.
        We discuss this further in \cref{sub:initial_conditions}.

     \item
        Loop over steps below until the relative difference in $\mathcal{L}$ changes by less than $\epsilon$.
        \begin{enumerate}
            \item
                Maximise $\mathcal{L}$ with respect to $\M$ while keeping $\pivec$, $\svec$ and $\beta$ constant.

            \item
                Maximise $\mathcal{L}$ with respect to $\pivec$ while keeping $\M$, $\svec$, $\beta$ constant, via the exact expression from Akagi \textit{et al.}
                \begin{equation}
                    \pi_i
                    =
                    \frac{%
                        \sum_t \sum_{j \in \Gamma_i \setminus \{i\}} \Mtij
                    }{%
                        \sum_t \sum_{j \in \Gamma_i} \Mtij
                    }
                    \,.
                \end{equation}

            \item
                Iteratively optimise $\mathcal{L}$ with respect to $\svec$ and $\beta$.

                In contrast with Akagai \textit{et al.}, who use the Minorisation-Maximisation algorithm for this step, we optimise the $\svec$ and $\beta$ dependent part of $\mathcal{L}$ directly.
                Our experience was that the former approach can become ``stuck'' during an evaluation.

                The only part of $\mathcal{L}$ that depends on $\svec$ and $\beta$ is $\mathcal{L}_1$ and it can be rearranged into a target function $f$ defined by
                \begin{gather}
                    f
                    =
                    \sum_{i \in V} \left(
                        A_i \log(s_i)
                        - B_i \log\left(\sum_{k \in \Gamma_i \setminus \{i\}} s_k \exp(-\beta d_{ik})\right)
                    \right)
                    - \beta D
                    \\
                    A_i
                    =
                    \sum_t \sum_{j \in \Gamma_i \setminus \{i\}} \Mtji
                    \\
                    B_i
                    =
                    \sum_t \sum_{j \in \Gamma_i \setminus \{i\}} \Mtij
                    \\
                    D
                    =
                    \sum_t \sum_{i \in V} \sum_{j \in \Gamma_i \setminus \{i\}} d_{ij} \Mtij
                    \,.
                \end{gather}
                The derivation of $A_i$ requires reordering the sum containing $\svec$.
                This resummation obscures the scale-independence of $\svec$ seen in \cref{eq:prob,eq:L}, and is only valid when the matrix $\mathbf{d}$ of distances $d_{ij}$ is symmetric.\footnote{The $d_{ij}$ is effectively a cost function for travel between Block-$i$ and Block-$j$. We have assumed this is symmetrical and $d_{ij} = d_{ji}$ but in principle this could be (for example) time-dependent and incorporate congestion related delays.  We do not consider these possible asymmetries in $d$.} We proceed as follows:
                \begin{enumerate}
                    \item
                        Optimise $f$ with respect to $\svec$.
                        There is a closed form for this:
                        \begin{equation}
                            s_i
                            =
                            \frac{A_i}{\sum_k C_k \exp(-\beta d_{k_i})}
                        \end{equation}

                    \item
                        Normalise $\svec$ with
                        \begin{equation}
                            \svec \mapsto \frac{\svec}{\max(\svec)}
                            \,.
                        \end{equation}
                        This is done to avoid numerical problems where $|\svec| \to 0$ otherwise.

                    \item
                        Maximise $f$ with respect to $\beta$.
                        This maximisation is done with the bounded Brent algorithm.

                \end{enumerate}
                We found that this optimisation of $\svec$ and $\beta$ would occasionally enter a closed loop.
                When this happens, we terminate the optimisation of $\svec$ and $\beta$ and return to the optimisation of $\M$ and $\pivec$ before trying again.

        \end{enumerate}

\end{enumerate}

We note the similarity of the procedure described here to the well-known Expectation Maximisation (EM) algorithm \cite{10.2307/2984875}. The EM algorithm is a method for performing maximum likelihood estimation in the presence of latent variables and has broad applicability to diverse fields from computational biology to machine learning \cite{2013arXiv1301.7373F, Do2008}. The EM algorithm works by iteratively improving parameter value estimates through alternating between an expectation step and a maximisation step. In the expectation step, a function for the expectation value of the log-likelihood is computed using the existing estimations for the parameter values. In the subsequent maximisation step, new parameter values are computed which maximise the value of the previously determined function. This process is then iterated until the desired convergence criteria are met. An adaptation of the EM algorithm which closely resembles our approach is known as Expectation Conditional Maximisation (ECM) \cite{10.2307/2337198}. In this case, each maximisation step is subdivided into a series of conditional maximization steps in which maximisation with respect to each parameter is undertaken individually, while all other parameters remain fixed. A detailed comparison between the efficacy of the algorithm implemented in this work and other variants of EM/ECM is out of scope here, but warrants further investigation going forward.

The majority of the implementation of this ``exact'' maximisation algorithm is in idiomatic Python and Numpy~\cite{vanderWalt:2011bqk}.
Some calculations make use of Numba~\cite{numba}, but ultimately this was not a major performance gain over vanilla Numpy.
Including the analytic functions in \cref{eq:der_L_0,eq:der_L_1,eq:der_L_2,eq:der_C} for the derivative of the likelihood improved performance by multiple orders of magnitude.

\begin{figure}[tb]
    \centering
    \includegraphics[width=0.8\linewidth]{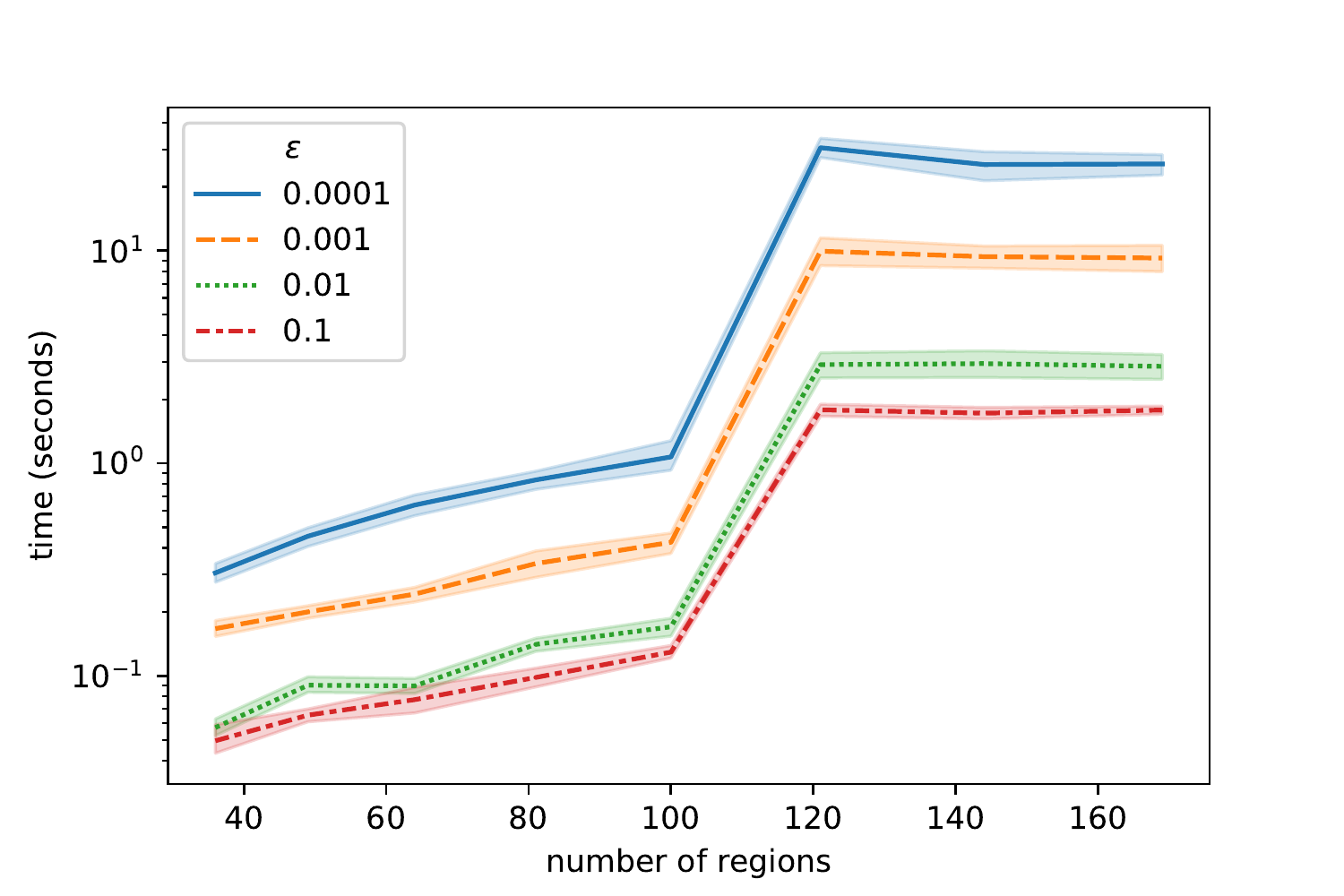}
    \caption{%
        Plot of the run time against number of regions.
        The shaded bands represent the standard deviation across 18 runs with 3 random seeds and 2 noise amplitudes for the synthetic data, and 3 choices of $\lambda$ in the solver.
        Clearly, demanding higher precision increases the run time but it remains manageable even as the number of regions grows beyond 100.
        All simulations were run on a consumer grade computer.
}
    \label{fig:time}
\end{figure}

\Cref{fig:time} shows wall clock computational time for a representative test problem based on synthetic data. While a plateau is observed in \Cref{fig:time} when the number of regions exceeds $\sim 120$, we would of course expect further increases in run time for much larger data sets. The details of this will depend upon multiple factors such as NumPy's usage of multiple CPUs for certain operations, and the available memory. In practice, estimating movements within the approximately $800$ SA2 blocks in the Auckland and Waikato regions took $\sim 30$ seconds on a laptop; this is consistent with synthetic data.  Consequently, the numerical performance of our implementation of the exact algorithm  presents no significant challenges for any currently envisaged applications, and appears to improve significantly on that reported by Akagi \textit{et al.}, presumably thanks to the use of the analytic derivatives.

\section{Synthetic data}%
\label{sec:simulated_data}

We do not have the true values of $\M$ for the cellphone data, whereas Akagi \textit{et al.} had access to trajectory information. However, we can test against synthetic data that strictly matches the assumptions laid out in \cref{sec:likelihood}.

We specify the number of regions we want to simulate, the distances $d_{ij}$ between them, cutoff distance $K$ and distance weighting $\beta$.
Then we stipulate vectors of gathering scores $\svec$ and departure probabilities $\boldsymbol{\pi}$ corresponding to each region.
From this, the simulator calculates the set of possible destinations $\Gamma_i$  of each region and probabilities $\thetaij$ for moving to each from \cref{eq:gamma,eq:transition_probability}.

We specify an initial distribution of people in each region as a vector $N_0$ and number of time steps to take.
Then for each time step $t$ and region $i$, the simulator loops over each of the people currently in region $i$ and assigns them to be in region $j$ at time $t+1$ with probability $\thetaij$.
This defines a ``true'' $\Mtij$ Optionally, the simulator also randomly adds or removes people before calculating where they go.
This allows us to test against scenarios that do not conform exactly to the assumptions in \cref{sec:likelihood}. 

\begin{figure}[tb]
    \centering
    \includegraphics[width=\linewidth]{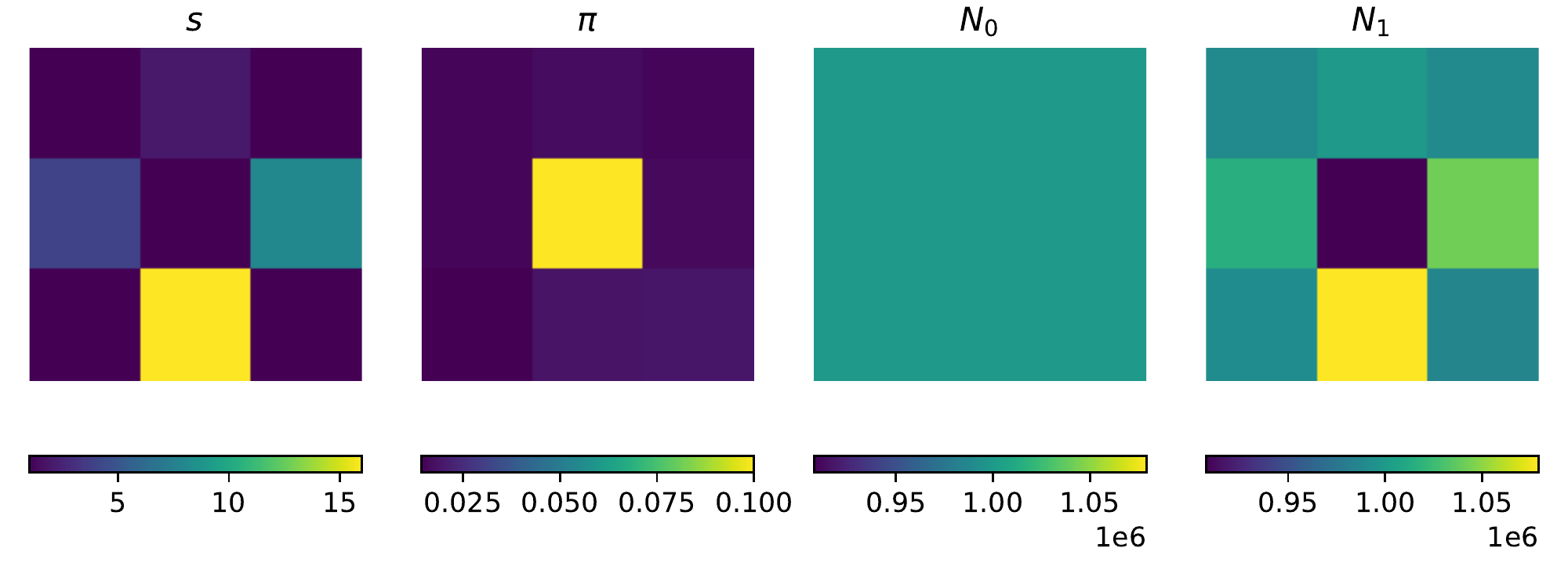}
    \caption{%
        From left to right: true $s$, true $\pi$, initial counts, and final counts for simulated synthetic data.
        This data has 9 regions, each with an initial count of $N_{0} = 1,000,000$.
        People leave each region with probability $\pi$ and each region has a gathering score $s$.
        One can see that the region with higher $\pi$ has a net loss in $N_{1}$.
        The regions with larger $s$ have correspondingly larger net gains.
    }%
    \label{fig:small_true}
\end{figure}

\Cref{fig:small_true} shows a  deliberately simple setup: an initially uniform population distributed among 9 regions arranged in a regular $3 \times 3$ grid. The distance between the centres of the outermost cells on each side of the grid is therefore 2 units.

We set the distance threshold $K=2$ and $\beta=1$.
Because the grid (expressed in terms of the centroids) has a width of 2, only corner to corner travel is disallowed  with $d = 2 \sqrt{2} > K$.
The departure probabilities $\pivec$ are sampled uniformly from $0.01$--$0.02$, other than the central region which has a probability of $0.1$.
This higher departure probability is evident in the populations after one time step; the central region has a larger net loss of people than the outer regions.
The gathering scores $\svec$ are set to 1 other than 4 regions shown in the left panel of \cref{fig:small_true}.
The gathering scores have the expected effect; regions with larger $s$ have correspondingly larger net gains.

\begin{figure}[tb!]
    \centering
    \includegraphics[width=\linewidth, trim={0.5cm 1cm 0 0}]{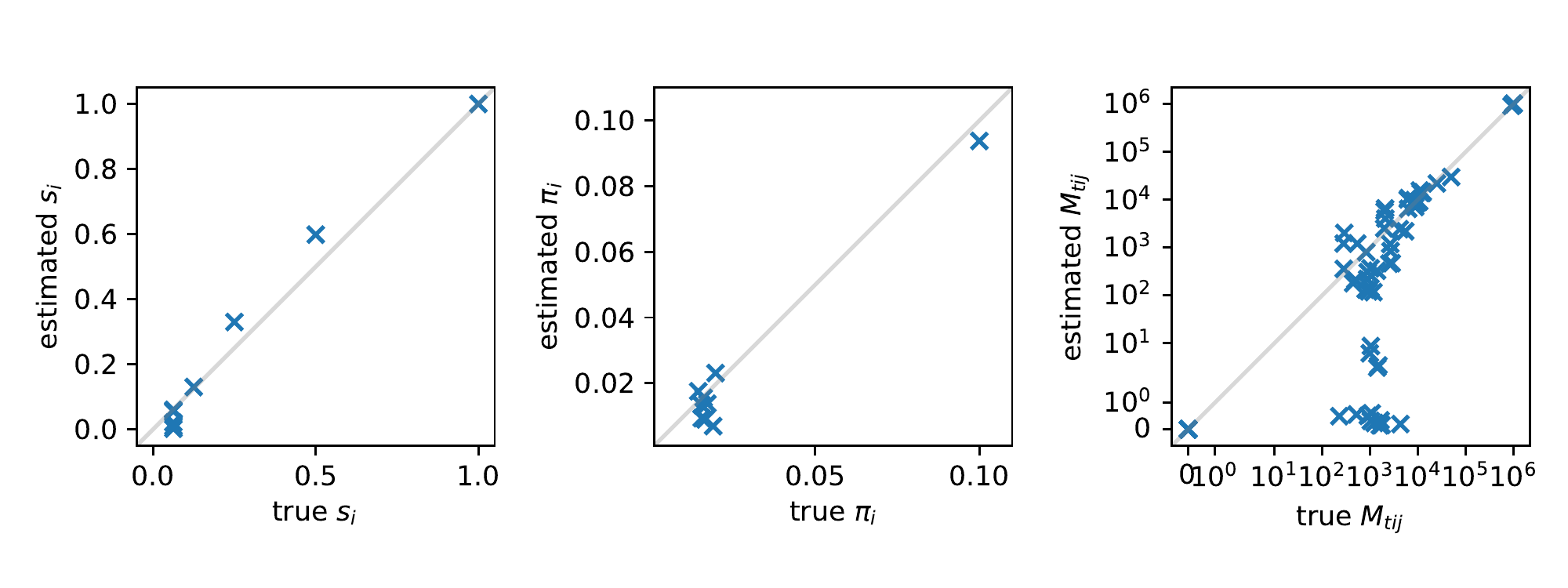}
    \caption{%
        Scatter plots of the true and estimated values of $s_i$, $\pi_i$ and $\Mtij$.
        The accuracy of the $s$ and $\pi$ estimates are relatively good, but there are a number of the transition matrix elements $M$ that are severely underestimated.
        The large elements of $\Mtij \sim 10^6$  are in the diagonals; these are people who did not move.
        Figures such as this are presented as 2-D histograms in the following sections, where there are too many points for a sensible scatter plot.
    }%
    \label{fig:small_est}
\end{figure}

\Cref{fig:small_est} shows the results of applying our implementation of the exact algorithm to the data described above.
It is able to get good estimates of the gathering scores and departure probabilities.
However, the estimates of $\Mtij$ and $\beta$ are poor.
There are a number of transitions that are severely underestimated.
In addition, $\beta$ is estimated at 0.08 rather than the true value of 1.0.
Poor estimates of $\beta$ are a recurring theme in the following sections.

\section{Implementation and Validation}%
\label{sec:validation_of_method_and_implementation}

We now  consider the stability of the results against changes in free parameters in the algorithm, and specific issues that arise when we apply the Akagi \textit{et al.} algorithms to the present use-case.

\subsection{Scaling}%
\label{sub:scaling}

As mentioned in \cref{sec:likelihood}, \cref{eq:L} assumes that Stirling's approximation $\log{n!} \approx n \log n - n$  applies to the elements of the transition matrix, or $\Mtij \gtrsim 1$.
However, this assumption is violated by fairly typical real-world data.
SA2 regions generally contain $\mathcal{O}(1,000)$ people and if $\mathcal{O}(100)$ people enter or leave in an hour with $\mathcal{O}(100)$ allowed destinations and origins, some transitions will necessarily involve ``fractional'' numbers of people.
These fractional numbers of people should be interpreted as probabilities.

\begin{figure}[p!]
    \centering
    \includegraphics[width=\textwidth]{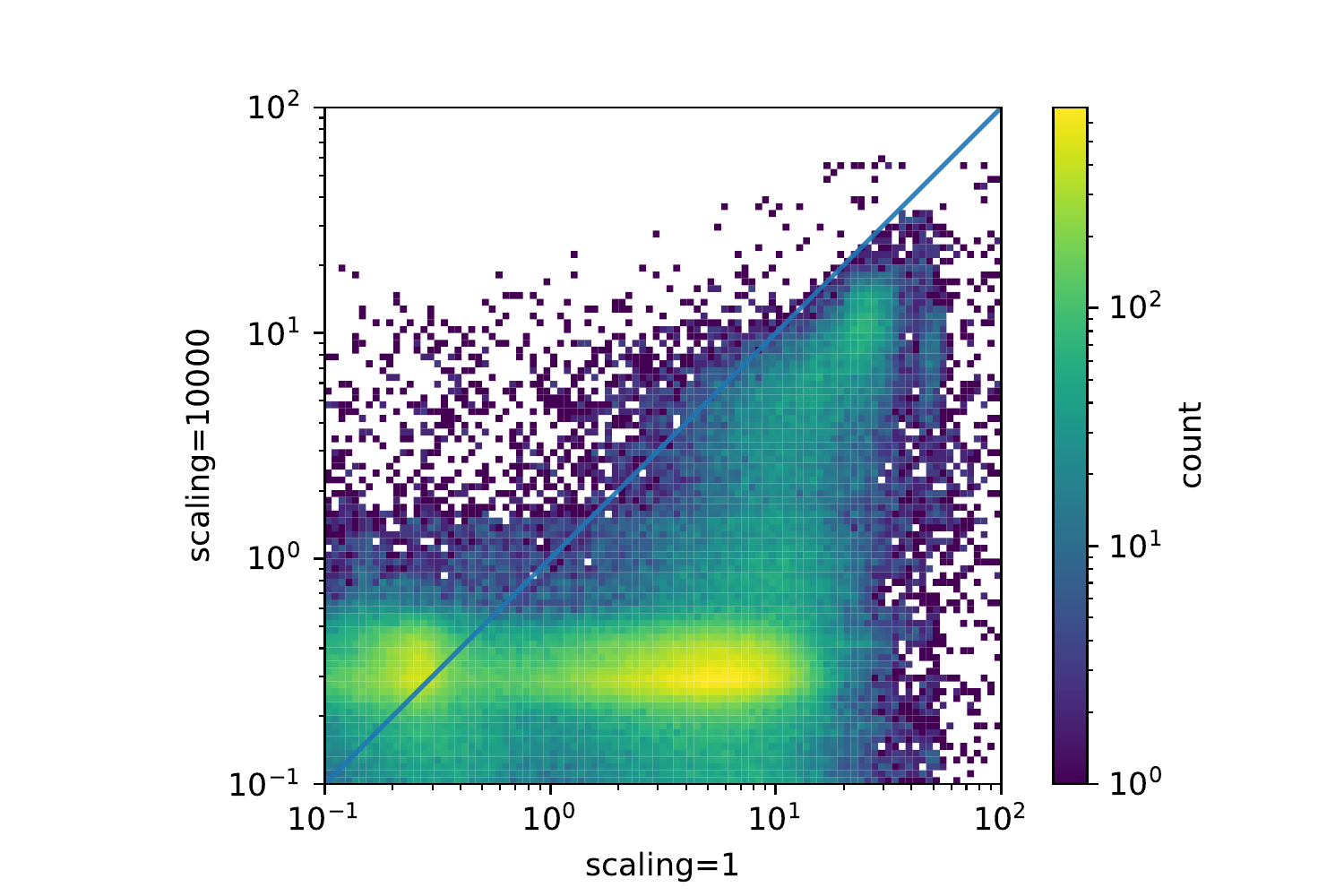}
    \includegraphics[width=\textwidth]{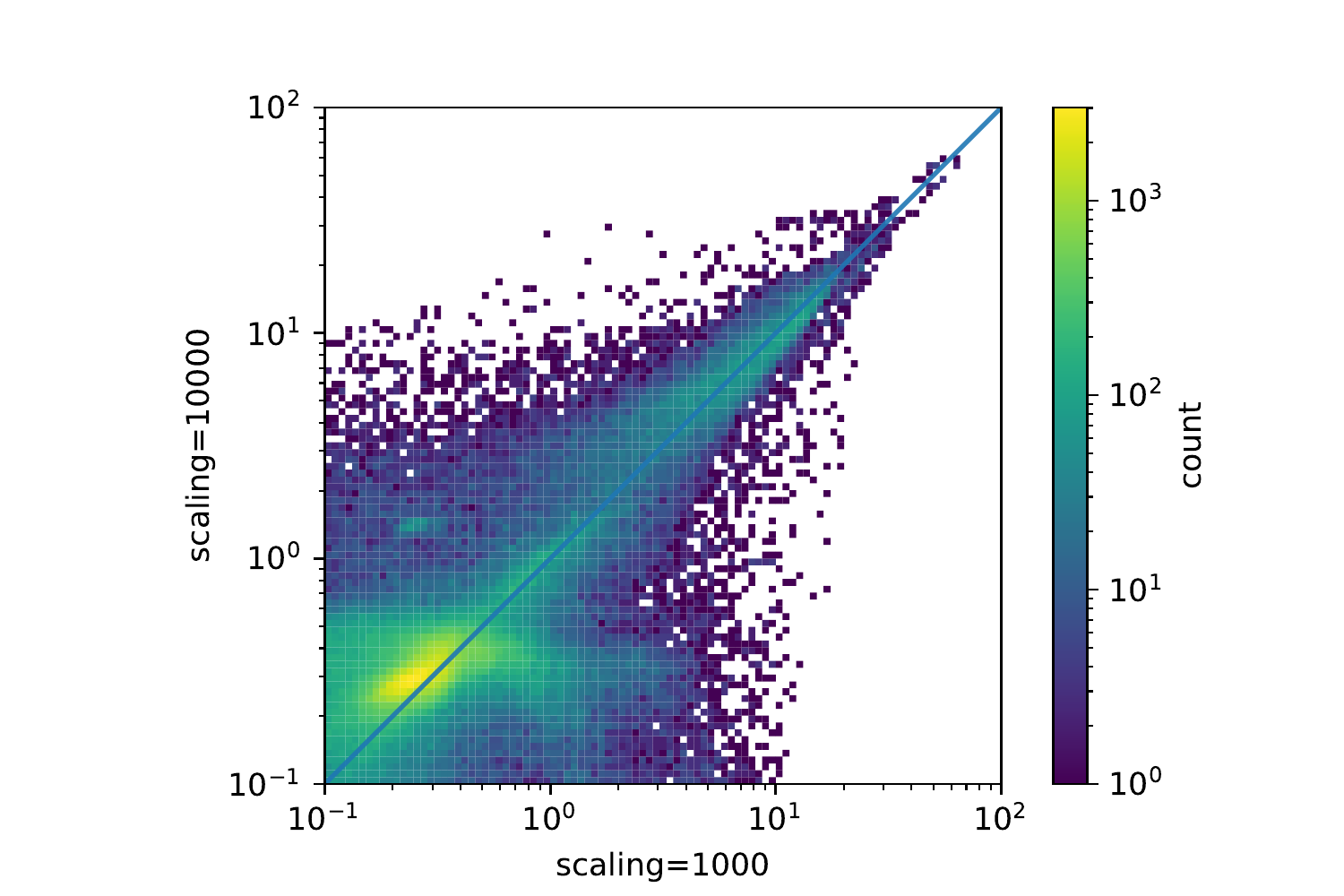}
    \caption{%
        Histograms comparing computed $M_{tij}$ for different population scalings.
        The data is for transitions between the 798 unique SA-2s in the regional councils of Auckland Region and Waikato. The plots show the counts of $M_{tij}$ for pairs of scalings; with perfect agreement all elements would lie on the diagonal, up to the scatter arising from the large number of near-degenerate solutions to the maximisation problem.    The top panel compares the raw counts (a scaling of 1) with a scaling of 1000 (y-axis). The bottom panel compares a scaling of 1000 (x-axis) and 10,000 (y-axis).
        }
    \label{fig:scaling}
\end{figure}

We have found that the inapplicability of Stirling's approximation can be ameliorated by scaling up the number of people to the point where all allowed transitions have $M_{tij} \gtrsim 1$. The cost function, \cref{eq:cost} is quadratic in this scale factor, so one must simultaneously rescale $\lambda$ to compensate. For sufficiently large scalings the results become scaling independent.
We checked that this is true by comparing the results for the SA2s contained in the combined Auckland Region and Waikato on February 18th between 7am and 9am.
There are 798 regions in total, but the small number with less than 100 people are dropped from the analysis of scaling by 1000 and 10,000, as shown in \cref{fig:scaling}.
This strategy is not necessarily perfect --- the $\Mtij \log(\Mtij)$ term in $\mathcal{L}_2$ is non-linear in $\Mtij$  and requires more detailed analysis  --- but will be more robust than using unscaled populations.
All other results shown in the following sections use a scaling large enough to ensure that the computed $\Mtij \gtrsim 1$ and these are then rescaled back to their original values.

\subsection{Repeatability and Initial conditions}%
\label{sub:initial_conditions}

The solver needs initial conditions that do not represent pathological states.\footnote{Note that `initial conditions' does not refer to values at $t=0$ but to the initial guesses for $\beta$, $\mathbf{s}$, $\boldsymbol{\pi}$, and the entire $\mathbf{M}$ matrix, prior to optimisation.}
As an initial guess we make the following ``static'' choice
\begin{gather}
    \pi_i = 0.02
    \\
    s_i = 0.02
    \\
    \beta = \frac{50}{\max(d)}
    \\
    \label{eq:M_initial_static}
    M_{tij}
    =
    \begin{cases}
        N_{ti}, &\text{for $i=j$ }
        \\
        0, &\text{for $i \ne j$}
    \end{cases}
    \,,
\end{gather}
where the last line implies that no-one moves between blocks.
This is not self-consistent, since if the off-diagonal $M_{tij} =0$, the $\pi_i$ and $s_i$ should also vanish, but these initial conditions can yield plausible results.
Varying the starting position causes the algorithm to converge on different maxima.

\begin{figure}[p!]
    \centering
    \includegraphics[width=0.8\linewidth]{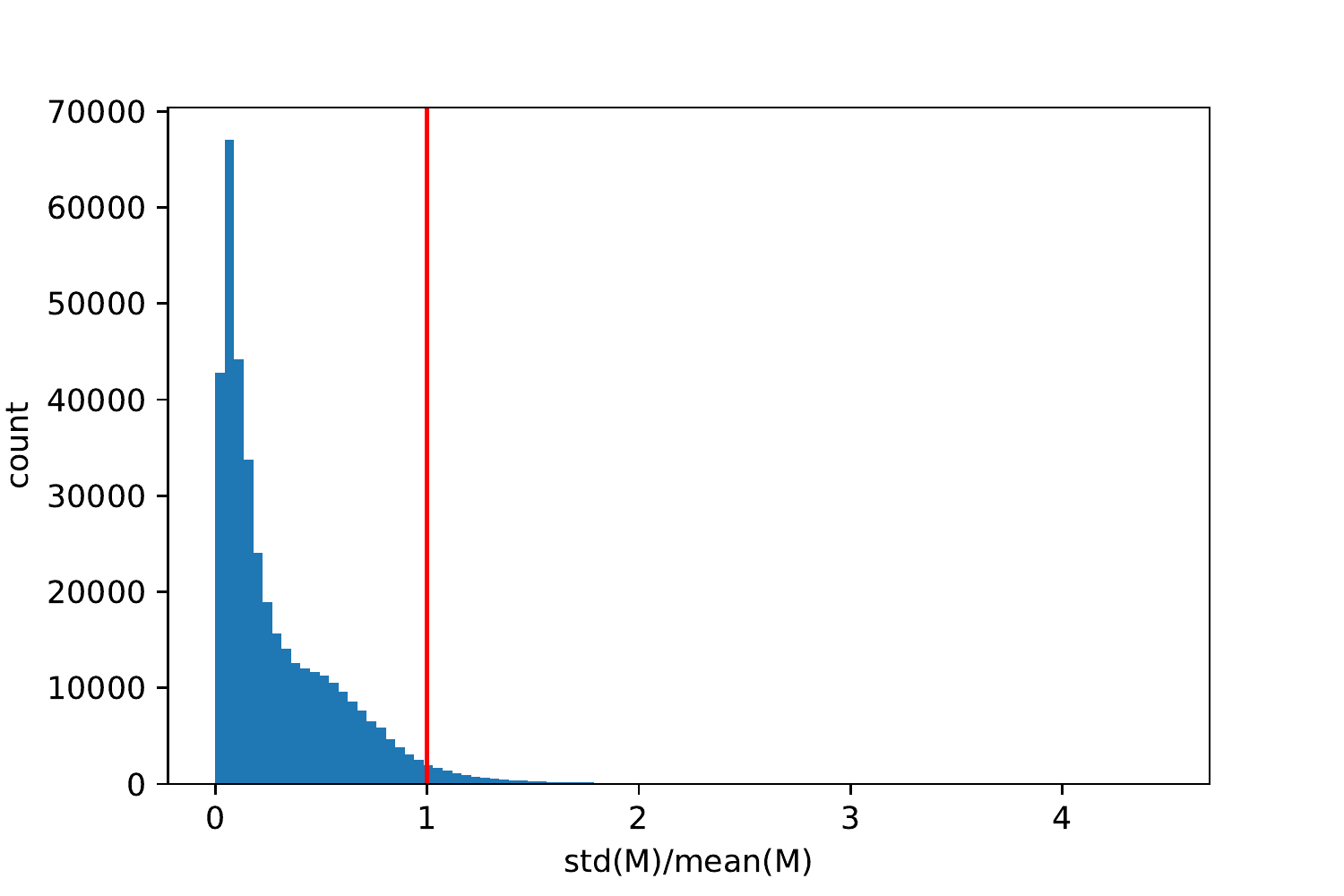}
    \includegraphics[width=0.8\linewidth]{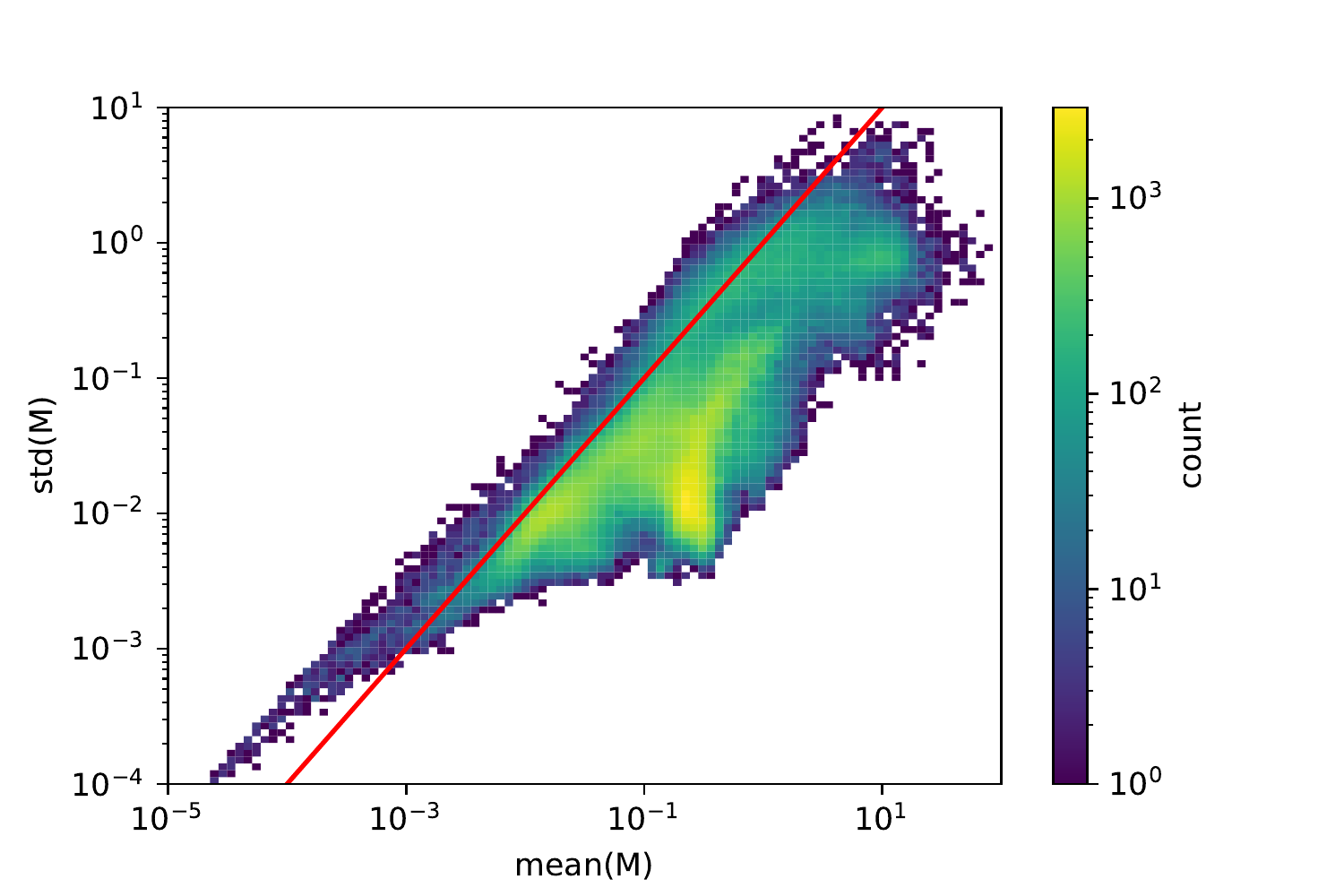}
    \caption{%
        Top:
        Histogram of the normalised standard deviation $\mathrm{std}(\Mtij)/\bar{M}_{tij}$ of $M$ values.
        The mean is with respect to 20 runs, each with the initial conditions of \cref{eq:M_initial_static_jitter} and different seeds for the random jitter; in most cases  $\mathrm{std}(M_{tij})/\bar{M}_{tij}$ is significantly less than unity.
        Bottom:
        Two-dimensional histogram of the same data.
        When the standard deviation is  below the red line it is less than the mean value; $M_{tij}$ above this line have large scatter between runs.   Note the logarithmic colour scale on this plot; the most common points and the largest $M$ values are below the line.
        Data is for SA2s contained in the combined Auckland Region and Waikato, on February 18th at 7am 8am, and 9am.
    }
    \label{fig:variance}
\end{figure}

We first test the sensitivity to initial conditions by adding a random scatter to the initial guess:
\begin{gather}
    \label{eq:M_initial_static_jitter}
    M_{tij}
    =
    \begin{cases}
        N_{ti} + \delta_{tii}, &\text{for $i=j$ }
        \\
        \delta_{tij}, &\text{for $i \in \Gamma_i \setminus \{i\}$ }
        \\
        0, &\text{for $i \notin \Gamma_i$}
    \end{cases}
\end{gather}
where $\delta_{tij}$ is sampled uniformly from the range $[0, N_{ti})$.
In \cref{fig:variance} we show that this scatter in the initial conditions does not have a drastic impact on the output by analysing data from SA2s contained in the combined Auckland Region and Waikato regions, on February 18th at 7am 8am, and 9am.
We quantify this sensitivity by computing the mean and standard deviation for the values of each of the $\Mtij$ from a sequence of 20 runs with different random perturbations to the original initial condition  \cref{eq:M_initial_static}. We find that the ratio of the standard deviation to the mean is small for the vast majority of $\Mtij$ for the cases we consider.

We also consider a ``moving'' initial guess,
\begin{equation}
    \label{eq:M_initial_moving}
    M_{tij}
    =
    \begin{cases}
        N_{ti}, &\text{for $i=j$ }
        \\
        \frac{\left|N_{ti} - N_{t+1, i}\right|}{|\Gamma_{i} \setminus \{i\}|}, &\text{for $j \in \Gamma_i \setminus \{i\}$}
        \\
        0, &\text{for $i \notin \Gamma_{i}$}
    \end{cases}
    \,.
\end{equation}
This encodes an expectation that most people stay where they are and that the number of people moving out of a region is on the order of the change in its population (regardless of whether that change is a net inflow or outflow).

\begin{figure}[p!]
    \centering
    \includegraphics[width=0.8\linewidth]{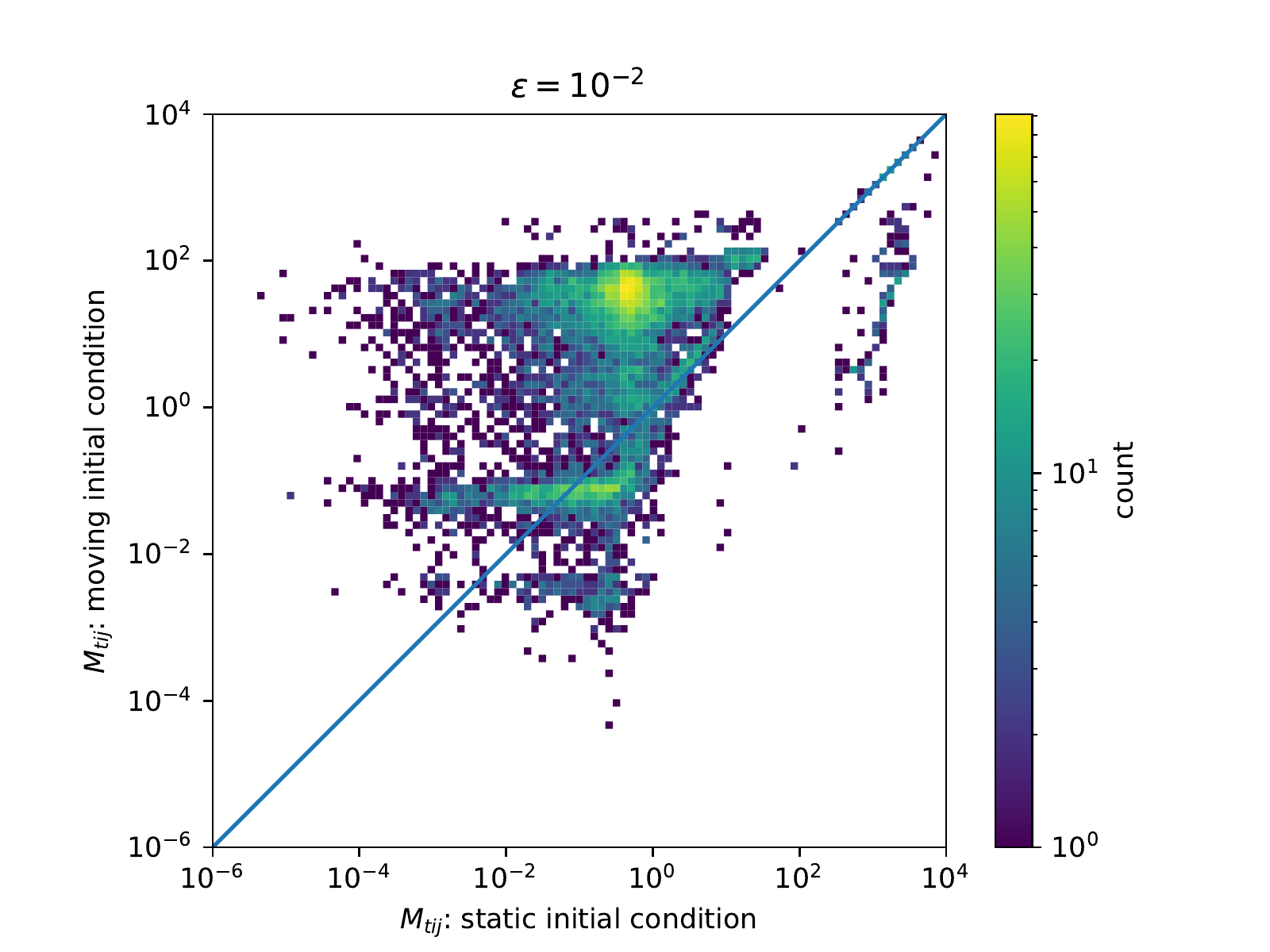}
    \includegraphics[width=0.8\linewidth]{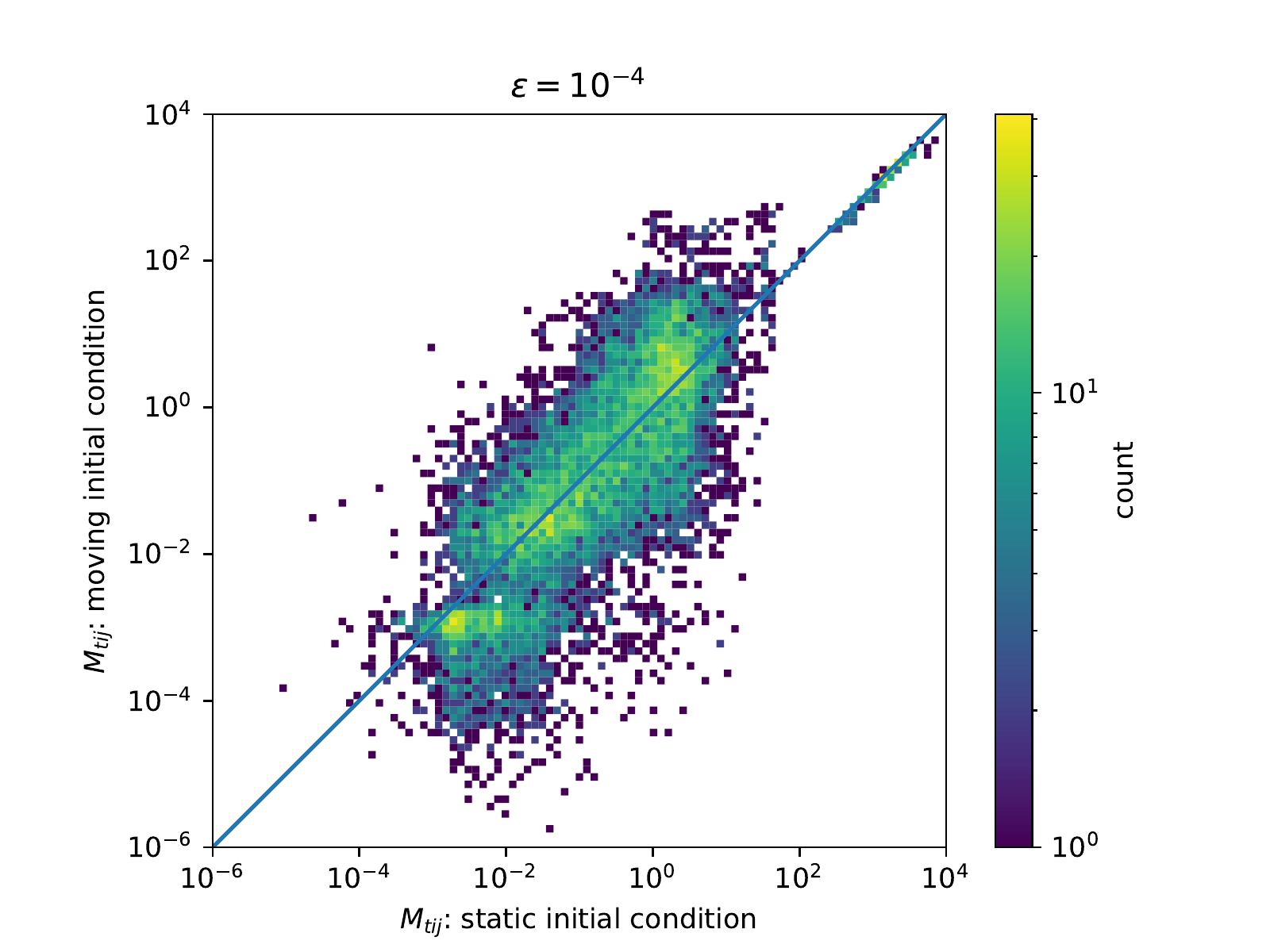}
    \caption{%
        Histograms of inferred $\M$ values with different initial conditions.
        The top panel has $\epsilon=10^{-2}$ and the bottom has $\epsilon=10^{-4}$.
        Static initial conditions ($x$-axis) start with diagonal transition matrices; i.e.\ no-one moves, as in \cref{eq:M_initial_static_jitter}; the $y$-axis have initial conditions for which many people move, as in \cref{eq:M_initial_moving}.
        With a loose convergence parameter the final result reflects the initial choice;
        setting $\epsilon = 10^{-4}$  eliminates most sensitivity to initial conditions.
        Data is from the 63 most populated areas in Southland Region, on 11 February 2020 at 6am, 7am, 8am, 9am and 10am.
    }
    \label{fig:different-initial-conds}
\end{figure}

In \cref{fig:different-initial-conds} we compare the two initial conditions choice described above.
We use data from the 63 most populated areas in Southland Region, on 11 February 2020 at 6am, 7am, 8am, 9am and 10am.
There is a clear discrepancy when $\epsilon = 10^{-2}$ but moving to a more stringent $\epsilon = 10^{-4}$ eliminates much of this bias.

\subsection{Sensitivity to  \texorpdfstring{$\epsilon$}{epsilon} and \texorpdfstring{$\lambda$}{lambda}}%
\label{sub:sensitivity_to_parameters_epsilon_and_lambda_}

The values  $\epsilon$ and $\lambda$ have an impact on the output of the algorithm.
We used the normalised absolute error (NAE) to quantify the error in these fits, which is defined to be
\begin{equation}
    \frac{\sum\limits_{t,i,j}\left\vert M^*_{tij}-M_{tij}\right\vert}{\sum\limits_{t,i,j} M^*_{tij}},
\end{equation}
where $M^*_{tij}$ represents the `true' $M$ values from the simulated data. We note that the NAE may be misleading in cases where there are a small number of regions for which the $M_{tij}$ are highly inaccurate if these regions have relatively large populations.

We examined the impact of $\epsilon$ and $\lambda$ by running the exact estimator on simulated data, as in \cref{sec:simulated_data}.
We assume $15^2$ cells distributed on a regular $2 \times 2$ grid and a distance cutoff $K = 1.5$, so that each region has $100$ to $225$ possible destinations.
The initial number of people, gathering scores and leaving probabilities in cell $i$ were set to
\begin{gather}
    N_{0,i} = \nu \exp\left( - {(\sqrt{x_i^2 + y_i^2} - r_0)}^2 \right)
    \\
    s_i = \exp(-4 (x_i^2 + y_i^2))
    \\
    \pi_i = \frac{1}{10} \frac{N_{0,i}}{\max_j (N_{0j})}
    \\
    \beta = 1
\end{gather}
where $r_0 = 0.8$.
The gathering score is high at the center, and the departure probability is proportional to the initial number of people in a cell.
There is a higher density of people in a ring of radius $r_0$ around the center.
This is intended to be roughly analogous to people migrating from the outskirts of a city into the center.
We allowed a 10\% error in the number of people at each location during each time step.

\begin{figure}[p!]
    \centering
    \includegraphics[width=0.8\linewidth, trim={0 2cm 0 2cm}]{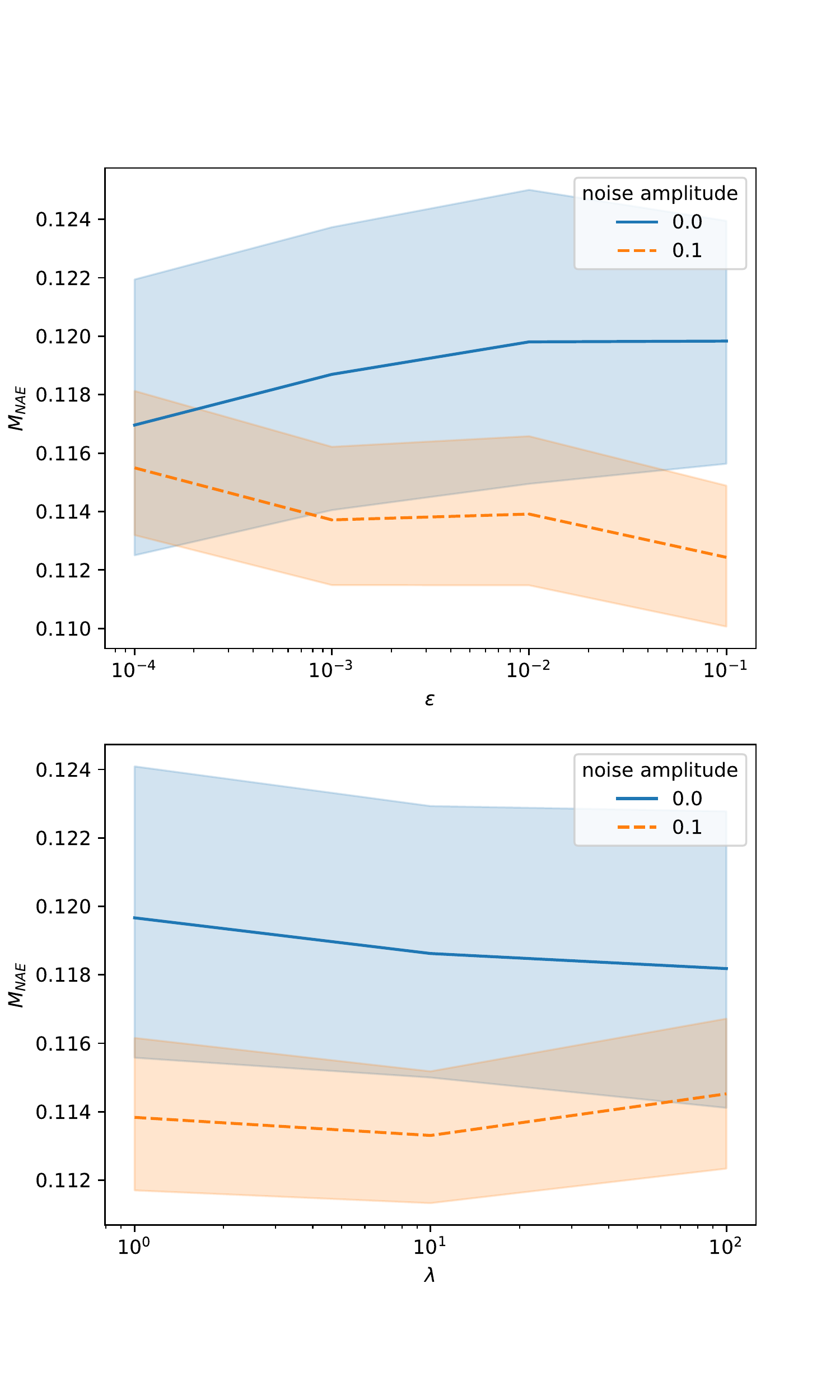}
    \caption{%
        The value of the NAE as compared to simulated data estimator parameters $\epsilon$ (top) and $\lambda$ (bottom).
        The error bands come from aggregating over 8 instances of $N$ and $\lambda$ (top) and $\epsilon$ (bottom).
        The blue solid line is the error when the simulated data conforms exactly to the assumptions of the likelihood \cref{sec:likelihood}.
        The orange dashed line assumes that there is an error of up to 10\% at each step.
        Interestingly, the estimator performs better on noisy data.
        Smaller values of $\epsilon$ do not have a clear advantage when there is noise in the data.
        On the other hand, $\lambda = 10 $ is better than 1 or 100.
    }
    \label{fig:eps-lambda-nae-2}
\end{figure}

The results are shown in \cref{fig:eps-lambda-nae-2}.
The absolute variance in the NAE as a function of both $\epsilon$ and $\lambda$ is not large.
Counterintuitively, we found that smaller values of $\epsilon$ do not necessarily give more accurate results by this measure, but the differences are not significant.
There is no obvious choice of $\lambda$; large values heavily penalise solutions where summing the people going in and out of regions does not match the known data $N$.
There are also numerical issues introduced by large $\lambda$; these seem much like the issues introduced with very small $\epsilon$.
Small values of $\lambda$ allow proposed solutions to have large violations of number conservation.
Given that the real-world data is known to have imperfect number conservation, some deviation should be allowed and a middle ground should be found.
The bottom panel in \cref{fig:eps-lambda-nae-2} confirms this intuition.

\section{Alternative Algorithm: Approximate Inference Method}\label{sec:approx_inference}

Akagi \textit{et al.} present an alternative method, which is billed as being less computationally expensive. This concern is less pressing, given the speedup applied to the exact algorithm.
We have implemented a variation of this algorithm in Python, with a few key differences. In particular,  Akagi \textit{et al.} bin regions by their separations but we cannot adopt this approach, given the irregular spacing of the SA-2 centroids.

\subsection{Summary of alternative algorithm}\label{sec:sum_alt}
We begin by defining the following parameters:

\begin{equation}
    X_{tij} \equiv M_{tji}, \quad Y_{ti} \equiv \sum_{j\neq i}M_{tij}, \quad Z_{ti}\equiv M_{tii}.
\end{equation}

Using these parameters, $\pivec$ and $f(\svec, \beta)$ are given by:

\begin{equation}
    \pi_i = \frac{\sum\limits_t Y_{ti}}{\sum\limits_t (Y_{ti} + Z_{ti})},
\end{equation}

\begin{equation}
    f(\svec, \beta) = \sum_{t,i,j} (X_{tij}\log s_i - \beta d_{ij} X_{tij}) -\sum_{t,i} Y_{ti}\log\Big(\sum_{k\neq i}s_{k}\exp(-\beta d_{ij})\Big),
\end{equation}

where $d_{ij}$ is the distance between centroids of SA-2 regions. We also define parameters $\theta_{ij}$ and $\mu_{ij}$ as follows:

\begin{equation}
    \theta_{ij} = \begin{cases}
  1 - \pi_i, &  (i = j)\\ 
  \pi_i\left(\frac{s_j \exp(-\beta d_{ij})}{\sum\limits_{k \neq i} s_k\exp (-\beta d_{ij})}\right), & (i\neq j)
\end{cases}
\end{equation}

\begin{equation}
    \mu_{ij} = \sum_{t} N_{tj}\theta_{ji}.
\end{equation}

Following Akagi {\em et al.}, the approximate log likelihood is given by

\begin{align}
    \mathcal{L}_{\text{approx}} = & \sum_{t,i,j}\big(X_{tij}\log(\mu_{ij}) + X_{tij} - X_{tij}\log(X_{tij})\big)\nonumber \\
    & + \sum_{t,i} \big(Y_{ti}\log(N_{ti}\pi_i) + Y_{ti} - Y_{ti}\log(Y_{ti})\big) \nonumber \\
    & + \sum_{t,i} \big(Z_{ti}\log(N_{ti}(1-\pi_i)) + Z_{ti} - Z_{ti}\log(Z_{ti})\big),
\end{align}
with the associated constraint function,
\begin{equation}
    C(X,Y,Z) = \sum_{t,i}\Big(\vert N_{ti} - (Y_{ti}+Z_{ti})\vert^2 + \vert N_{t+1, i } - \sum_{j} X_{tij}\vert^2\Big).
\end{equation}

We then have a log likelihood function for the final calculation of $M$ as follows:
\begin{align}
    \mathcal{L}_{\text{final}} = & \sum_{t,i} \log(1-\pi_i)M_{tii} + \sum_{t,i,j} \big(M_{tij} - M_{tij} \log(M_{tij})\big)\nonumber \\
    & + \sum_{t, i, j\neq i}\bigg(\log(\pi_i) + \log(s_j) - \beta d_{ij} - \log\Big(\sum_{k\neq i} s_k \exp(-\beta d_{ik})\Big)\bigg),
\end{align}
with associated constraint function:
\begin{equation}
    C(M) = \sum_{t,i}\Big(\vert N_{ti} - \sum_j M_{tij}\vert^2 + \vert N_{t+1,i} - \sum_j M_{tji}\vert^2\Big).
\end{equation}

The inference proceeds as follows:
\begin{enumerate}
    \item Initialise parameters $\M$, $X$, $Y$, $Z$, $\pivec$, $\svec$, and $\beta$,
    \item Maximise $\mathcal{L}_{\text{approx}}$ - $\frac{\lambda}{2} C(X, Y, Z)$,
    \item Update $\pivec$,
    \item Update $\svec$ and $\beta$ by maximising $f(\svec, \beta)$,
    \item Repeat 1 - 4 until specified convergence criterion is reached for the value of the approximate log likelihood,
    \item Calculation of $\M$ through Maximising $\mathcal{L}_{\text{final}}$ - $\frac{\lambda}{2}C(\M)$, using the final $\pivec$, $\svec$, and $\beta$ values calculated above.
\end{enumerate}

When optimising $\mathcal{L}_{\text{approx}}$ and $\mathcal{L}_{\text{final}}$, it is useful to define their analytic Jacobians, as it is  computationally expensive to compute approximate derivatives, along with analytic Jacobians for the constraint. These are  as follows:
\begin{align}
    \frac{\partial \mathcal{L}_{\text{approx}}}{\partial X_{tij}} & = \log(\mu_{ij}) - \log(X_{tij}),\\
    \frac{\partial \mathcal{L}_{\text{approx}}}{\partial Y_{ti}} & = \log(N_{ti} \pi_i) - \log(Y_{ti}),\\
    \frac{\partial \mathcal{L}_{\text{approx}}}{\partial Z_{ti}} & = \log(N_{ti}(1-\pi_i)) - \log(Z_{ti}),
\end{align}
with constraint function derivatives:
\begin{align}
    \frac{\partial \big(-\frac{\lambda}{2} C(X, Y, Z)\big)}{\partial X_{tij}} &= \lambda \Big(N_{t+1, i} - \sum_k X_{tik}\Big),\\[1.0ex]
    \frac{\partial \big(-\frac{\lambda}{2} C(X, Y, Z)\big)}{\partial Y_{ti}} &= \frac{\partial \big(-\frac{\lambda}{2} C(X, Y, Z)\big)}{\partial Z_{ti}} = \lambda\Big(N_{ti} - (Y_{ti} + Z_{ti})\Big).
\end{align}

For the final log likelihood, we have:
\begin{align}
    \frac{\partial \mathcal{L}_{\text{approx}}}{\partial M_{tii}} &= \log(1-\pi_i) - \log(M_{tii}),\\[1.0ex]
    \frac{\partial \mathcal{L}_{\text{approx}}}{\partial M_{tij\neq i}} &= \log(\pi_i) + \log(s_j) -\beta d_{ij} - \log\Big(\sum_{k\neq i} s_k \exp (-\beta d_{ik})\Big)- \log(M_{tij}),
\end{align}
with constraint function derivatives:
\begin{equation}
    \frac{\partial \big(\frac{-\lambda}{2}C(M)\big)}{\partial M_{tij}} = \lambda\Big(N_{ti} + N_{t+1, i} - \sum_{k} M_{tik} - \sum_{k} M_{tkj}\Big).
\end{equation}

\subsection{Performance of alternative algorithm}
\label{sec:alt_performance}

We implemented this algorithm in both Python 2 and Python 3, noting that the former tends to outperform the latter, apparently due to bugs within the Numba compiler in Python 3.  Our implementation was first tested using synthetic data.  Using the NAE as a measure of the performance of the algorithm it was found that for large data sets, it is beneficial to nest the main loop, as described by steps 1 to 6 above, within an outer loop.  This outer loop feeds the calculated $\M$ values back as initial conditions in the subsequent evaluation. For testing purposes, the outer loop is terminated either when the NAE reaches a specified target value, or when successive loops result in no further decrease in the NAE\@. This is only possible when the true values are known, as in our test case. Hence, when applying this algorithm to real-world data, one may choose to terminate the outer loop when the successive change in $M_{tij}$ values reaches a certain threshold.

As an example, using simulated data with 225 regions over 3 time steps gave an NAE of 0.046 after three iterations through the outer loop, compared to 0.154 with only one iteration. By comparison, the ``exact'' algorithm achieved an NAE of 0.100, so that in this case the alternative algorithm appears to perform better, though it does take more computation time. In this case we chose $\lambda=10$, a convergence criterion of $0.001\%$ on the approximate log-likelihood, and a tolerance $\texttt{ftol} = 10^{-4}$ within scipy.optimise.minimise for the $\M$ calculation.

We can also calculate the off-diagonal NAE, as the large values on the diagonal can dominate the NAE, obscuring how well the algorithm is able to identify regions with high gathering scores.
In this case, the off-diagonal NAE for the alternative algorithm was 0.279, compared with 0.558 for the ``exact'' algorithm, again indicating more accurate reproduction of the input data.\footnote{The synthetic data used in this test, along with the implementation used to analyse it, are available in the code base within the folder `nae-comparison'.}

\subsection{Discussion of alternative algorithm}\label{sec:discussion_alt}

Testing indicates that initialising the $\M$ arrays with the corresponding $\N$ values on the diagonal and small random numbers on the off-diagonal provides the best outcomes.\footnote{The introduction of explicit randomness in the $\M$ initialisation can make it difficult to compare successive runs. To overcome this one may fix the seed of the random number generator.} The alternative inference algorithm runs through the entire inference process multiple times, inputting the new $\M$ arrays as initial conditions in each run. In some cases this  leads to much improved results, but can also result in an `overshoot', whereby the off-diagonal elements become too high.

The output is highly sensitive to the value of $\lambda$, which controls the strength of the penalty terms. If $\lambda$ is too small, the algorithm tends to overpopulate the off-diagonals. Conversely, if the value is too high, all off-diagonal elements tend to zero. The optimal value of $\lambda$ varies on a case-by case basis, making it difficult to guess a suitable value in advance.

In addition to the algorithm's sensitivity to the $\M$ initialisation, $\lambda$ value, and number of complete inference loops, one must also consider the convergence criteria set for the approximate log-likelihood in the inner loop, and tolerances set in the optimisation routines as well. Tighter convergence constraints may increase computation time to an unacceptable degree, or may preclude convergence entirely.

The original Akagi {\it et al.\/} treatment introduced this algorithm for its greater efficiency, and it serves as a useful counterpoint to the ``exact'' version. However, given its relative fragility with respect to control parameters and the efficiency of our implementations it does not appear to offer a significant, generic advantage.

\section{Validation Test Case: Southland}\label{sec:south_alt}

We are unable to test the performance of the algorithms against real-world trajectory information. However, one may gauge how well the algorithm captures the essential features of daily travel by comparing its output to census data, and we take a very cursory look at this problem here.
We accessed the publicly available self-declared commuting trends from the 2013 census using the Stats NZ `Commuter View' tool~\cite{commview}. This tool presents the number of residents who commute out of that region for work and the number that commute in.
We can then compare the trends in the census data to the sum of the off-diagonal $\M$ matrix elements for outbound and inbound travel for each region on a standard weekday, assuming most workers travel to work in a time period from 6am to 10am.

For a simple test case, we have singled out the SA-2 regions which belong to the wider Southland district, discarding regions with missing data at any of the times of interest. Of the 65 SA-2 regions within the Southland district, only 2 regions had incomplete data, both with populations of less than $10$. This comparison is not exact. The subdivision of geographical regions within the census data does not match the SA-2 regions used in the telecommunications data so assumptions must be made when matching the data to our calculations. Furthermore, this method does not capture the varying work hours of the general populace, and is seven years out of date.

\begin{figure}[tb!]
    \centering
    \includegraphics[width=1.0\linewidth]{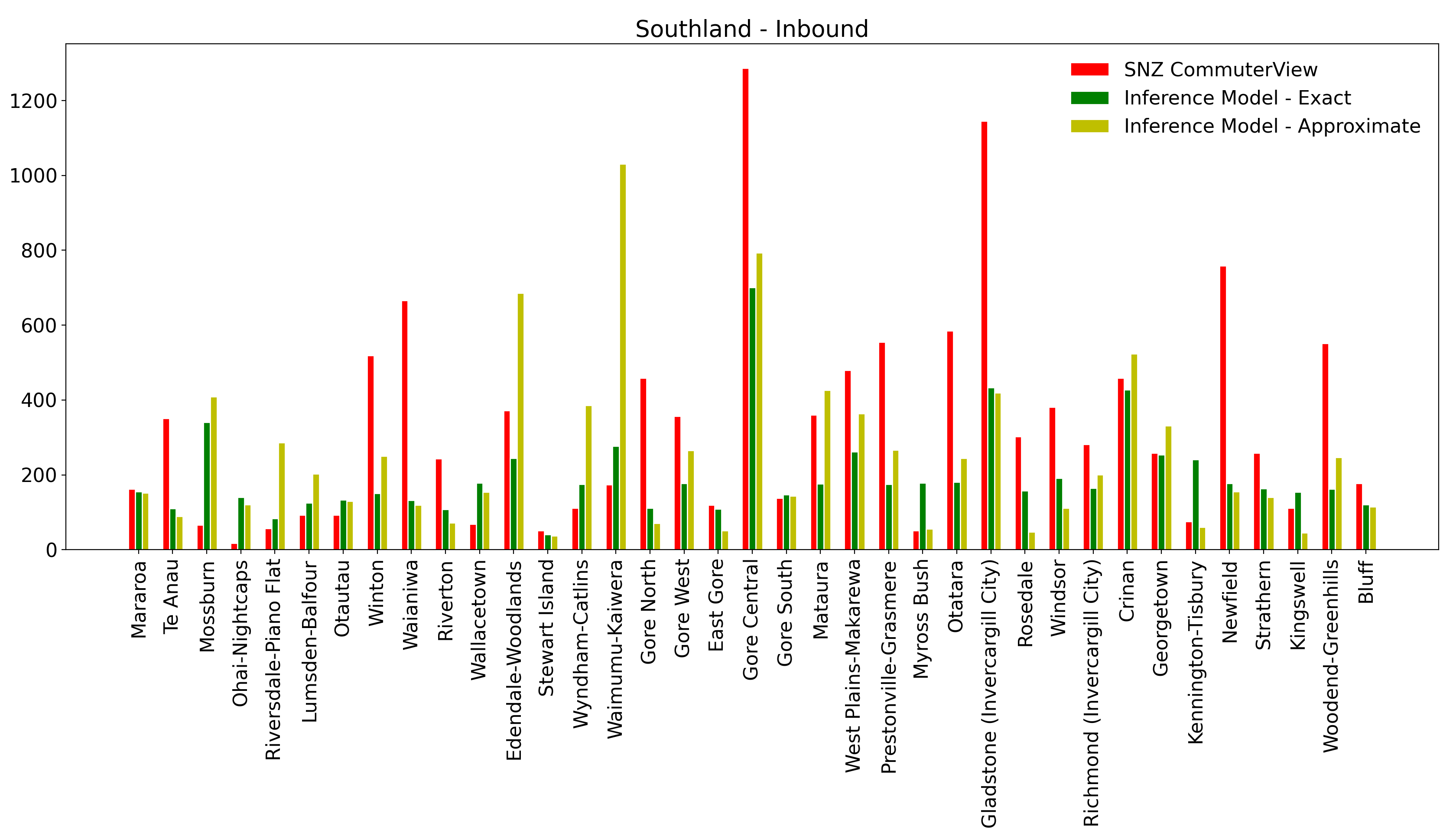}
    \includegraphics[width=1.0\linewidth]{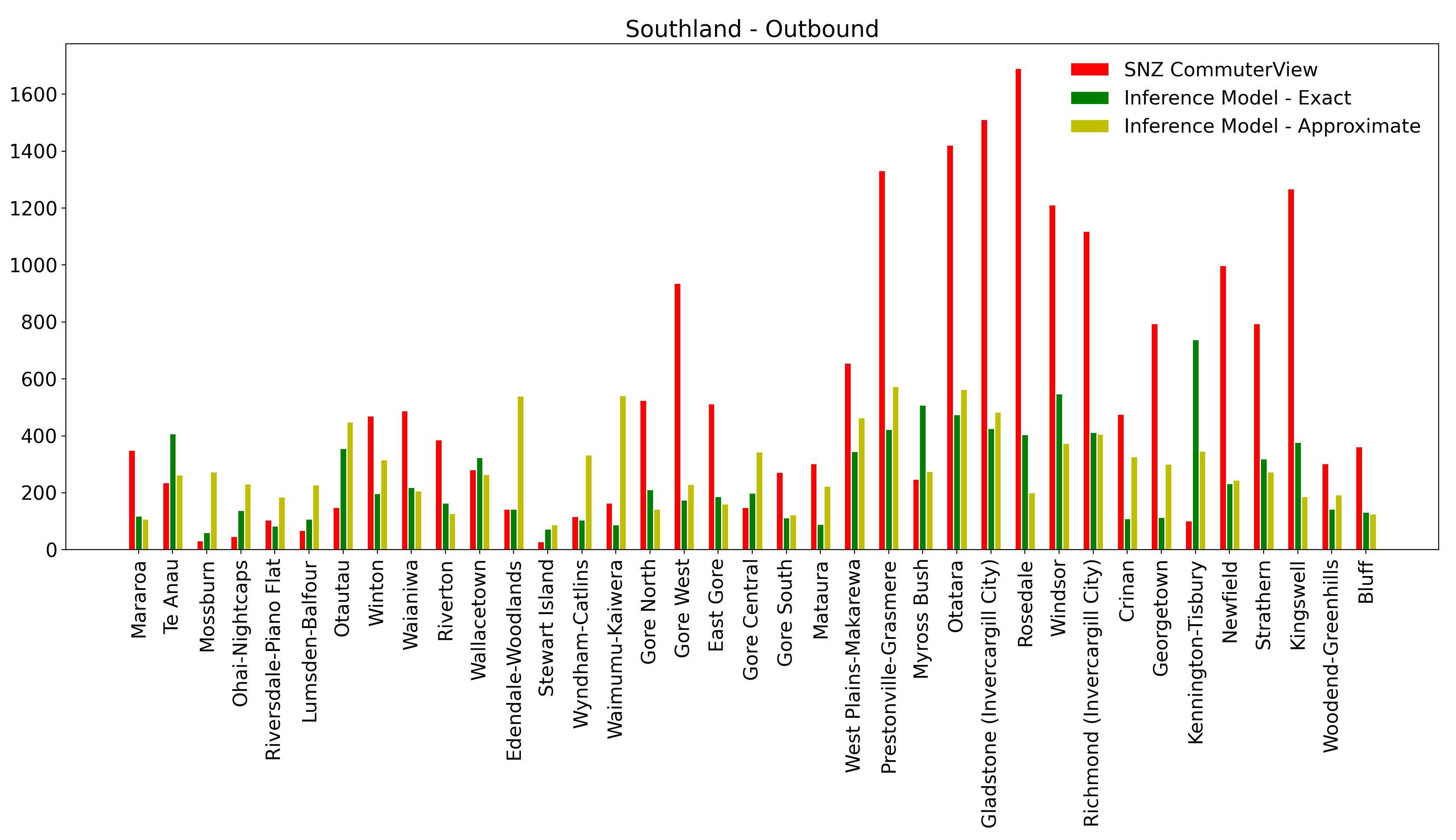}
    \caption{%
       Stats NZ Commuter View vs.\ movements estimated with the ``exact'' and ``approximate'' algorithm for Southland commuters.
       We used data from 11th February, for the five hours from 6am to 10am. SA-2 regions which do not correspond to a single commuter region are discarded in this analysis.
    }
    \label{fig:inoutbound}
\end{figure}

We ran the approximate inference algorithm for the telecommunications data from the 11th of Febuary, 2020 (Tuesday) from 6am to 10am. We then compare the total outbound and inbound commuters output by the algorithm with the census data. The results are displayed in \cref{fig:inoutbound}, including only those census regions which clearly correspond to one or more SA-2 regions. Moreover, cell coverage is not necessarily representative of an individual's exact location at any given moment, and  data between neighbouring regions may be somewhat mixed.\footnote{The code used to generate the comparisons shown here, along with the corresponding initial conditions, is  in the folder `southland-comparison'.}

While the algorithm appears to capture the essential features of the inbound commuting trends, with gathering concentrated within the main metropolis, the outbound inference fares significantly less well, with outbound commuters significantly underrepresented when compared to regions within the census data with high traffic. This comparison is strictly qualitative and ``one off'', and self-declared travel within inner-city regions may correspond to a change in cell-coverage regions, and vice-versa.

We also note that the `Commuter View' tool has since been re-branded to `Commuter Waka', and now also incorporates data from the 2018 census \cite{commwaka}. However, due to a particularly low response rate of 81.6\% to the 2018 census \cite{PES2018}, we choose to test our algorithm against the older data set - based upon the responses to the 2013 census only - which had a much higher response rate. It is hoped that better quality data will become available in future for more thorough verification testing.

\section{Summary}%
\label{sec:summary}

This Working Paper describes and re-implements two possible approaches to using count data to impute meso-scale human movement patterns.
We investigated the validity of the likelihood assumed and improved the minimization method. 
At this point it can analyse data for large fractions of the country (e.g. $\sim$800 out of $\sim$2000 SA-2 regions in a single run) via a computationally efficient and publicly available code.

The algorithm demonstrates qualitative agreement with simulated and real-world data. The actual numerical counts of people moving from one region to another come with some uncertainty, stemming from the fact that the problem is highly degenerate. In particular, we occasionally find estimated values of $\svec$, $\pivec$ $\M$ and $\beta$ that have a higher likelihood than the ``true'' solution, but nevertheless differ from it. Moreover, the model used here computes large numbers of ``point to point'' journeys, but in real world scenarios residents of two well-separated blocks may be more likely to interact in some third block, to which they both travel during the course of day.

That said, we can see a number of ways in which these approaches could be improved, and our implementations are publicly available. The algorithms could prove useful when averaged data is extracted from the outputs, such as estimates of mean distances travelled per day. Such averaged quantities may be more reliable than estimates of individual point-to-point journeys. 
The codes are therefore probably best regarded as providing order-of-magnitude estimates which may be of use in sensitivity testing complex infection propagation models, but should not be seen as yielding precise quantitative predictions.

While improvement is still needed, our work may have important applications in many areas relating to disease outbreak and containment. Namely, identification of the areas with highest gathering statistics could help to inform the most effective locations for lockdown boundaries, while a better understanding of common transit routes could help to identify high risk sub-regions outside of the most densely populated commercial and residential hubs. Finally, outputs from this algorithm may serve as useful inputs to more complex models of disease propagation.  

Specific directions for future work might include:
\begin{itemize}
    \item Adding a more nuanced distance metric, including driving distance, rather the centroid to centroid Euclidean distance.
    \item Considering a more complex penalty function, e.g.\ $\exp(-\beta d - \alpha d^2)$.
    \item
        Improving the quality of the data set.
        In particular, a count of cell phones in a block that were present in the previous hour would allow separate estimations of the $s_i$ and $\pi_i$ and would fix the diagonal elements of $\M$, but would likely raise few privacy concerns.
    \item
        Improving validation testing against census data, or traffic flow information for urban regions.
    \item
        Fitting $\pivec$, $\svec$, and (possibly) $\beta$ to census data rather than count data.
        One would have to justify the continued use of these values during periods of modified behaviour, such as when travel restrictions are in place.

        \item
        Developing an improved travel simulator to better test the model against a full realisation of movement patterns in areas of interest.
        
        \item Properly accounting for the fact that in most realistic datasets the transition probabilities will be time dependent, varying over the course of the (working) day.  
\end{itemize}

Finally, we emphasise that this overall problem is essentially an imputation exercise. Any  results obtained with them are estimates, and any model that uses them as inputs should be interpreted accordingly.

\section*{Biographical Note}

This work was undertaken in response to the Covid-19 emergency and was largely completed while New Zealand was at lockdown levels 3 and 4. At the time the work was undertaken the authors were all members of the theoretical cosmology group at the University of Auckland.

\nocite{Virtanen:2019joe,numba,vanderWalt:2011bqk}

\bibliographystyle{plain}
\bibliography{references}

\end{document}